 \documentclass[aps,pra,twocolumn,groupedaddress,amsmath,amssymb,showpacs,floatfix]{revtex4}
\usepackage{graphicx}
\usepackage{amsmath}
\usepackage{dcolumn}% Align table columns on decimal point
\usepackage{bm}% bold math

\newcommand{\tr}{\mathrm{Tr}}
\newcommand{\inp}{\mathrm{in}}
\newcommand{\out}{\mathrm{out}}

\begin{document}
\title{Linear optics and  quantum maps}
\author{A. Aiello}
\author{G. Puentes}
\author{J. P. Woerdman}
\affiliation{Huygens Laboratory, Leiden University\\
P.O.\ Box 9504, 2300 RA Leiden, The Netherlands}
\begin{abstract}
We present a theoretical analysis of the connection between
classical polarization optics and quantum mechanics of two-level
systems. First, we review the matrix formalism of classical
polarization optics from a quantum information perspective. In
this manner the passage from the  Stokes-Jones-Mueller description
of classical optical processes to the representation of one- and
two-qubit quantum  operations, becomes straightforward. Second, as
a practical application of our classical-\emph{vs}-quantum
formalism, we show how two-qubit maximally entangled mixed states
(MEMS), can be generated by using polarization and spatial modes
of photons generated via spontaneous parametric down conversion.
\end{abstract}

\pacs{03.65.Ud, 03.67.Mn, 42.25.Ja} \maketitle
%
%03.65.Ud    Entanglement and quantum nonlocality (e.g. EPR
%paradox, Bell's inequalities, GHZ states, etc.) (for entanglement
%production in quantum information, see 03.67.Mn; for entanglement
%in Bose-Einstein condensates, see 03.75.Gg)
%
%03.65.Nk    Scattering theory
%
%03.67.Mn    Entanglement production, characterization, and
%manipulation (see also 03.65.Ud Entanglement and quantum
%nonlocality; for entanglement in Bose-Einstein condensates, see
%03.75.Gg)
%
%42.25.Ja    Polarization
%
%42.50.Dv    Nonclassical states of the electromagnetic field,
%including entangled photon states; quantum state engineering and
%measurements (see also 03.65.Ud Entanglement and quantum
%nonlocality, e.g. EPR paradox, Bell's inequalities, GHZ states,
%etc.)
%
%
\section{Introduction}
Quantum computation and quantum information have been amongst the
most popular branches of  physics in the last decade
\cite{NielsenBook}.
One of the reasons of this success is that the smallest unit of
quantum information, the qubit, could be reliably encoded in
photons that are easy to manipulate and virtually free from
decoherence at optical frequencies \cite{Zeilinger99,Gisin02}.
Thus, recently, there has been a growing interest in quantum
information processing with linear optics
\cite{Knill,OBrien03,Skaar,Kok06} and several  techniques to
generate and manipulate optical qubits have been developed for
different purposes ranging from, e.g., teleportation
\cite{Bouwmeester,DeMaTele}, to quantum cryptography
\cite{Gisin02}, to quantum  measurements of qubits states
\cite{James01} and  processes \cite{OBrien04}, etc.
In particular, Kwiat and coworkers \cite{Peters05,Wei05} were able
to create and characterize arbitrary one- and two-qubit states,
using polarization and frequency modes of photons generated via
spontaneous parametric down conversion (SPDC) \cite{YarivBook}.

Manipulation  of optical qubits is performed by means of linear
optical instruments such as half- and quarter-wave plates, beam
splitters, polarizers, mirrors, etc., and networks of these
elements.
Each of these devices can be thought as an object where incoming
modes of the electromagnetic fields are turned into outgoing modes
by a \emph{linear} transformation.
 From a quantum information
perspective, this transforms the state of qubits encoded in some
degrees of freedom of the incoming photons, according to a
completely positive map $\mathcal{E}$ describing the action of the
device. Thus, an optical instrument may be put in correspondence
with a quantum map and vice versa.
Such correspondence has been largely exploited
\cite{Zhang04,Peters05,Wei05,Kok06} and stressed
\cite{Brunner03,Aiello061} by several authors.
Moreover, classical physics of linear optical devices is a
textbook matter \cite{BornWolf,Damask}, and quantum physics of
elementary optical instruments has been studied extensively
\cite{Leonhardt03}, as well. However, surprisingly enough, a
systematic exposition of the connection between classical linear
optics and quantum maps is still lacking.

In this paper we aim to fill this gap by presenting a detailed
theory of linear optical instruments from a quantum information
point of view. Specifically, we establish a rigorous basis of the
connection between quantum maps describing one- and two-qubit
physical processes operated by polarization-affecting optical
instruments, and the  classical matrix formalism of polarization
optics. Moreover, we will use this connection to interpret some
recent experiments in our group \cite{Puentes06b}.

We begin in Section II by reviewing the classical theory of
polarization-affecting linear optical devices. Then, in Section
III we show how to pass, in a natural manner, from classical
polarization-affecting optical operations to one-qubit quantum
processes. Such passage is extended to two-qubit quantum maps in
Section IV. In Section V we furnish two explicit applications of
our classical-\emph{vs}-quantum formalism that illustrate its
utility. Finally, in Section V we summarize our results and draw
the conclusions.
\section{Classical polarization optics}
In this Section we focus our attention on the description of
non-image-forming polarization-affecting optical devices. First,
we shortly review the mathematical formalism of classical
polarization optics and establish a proper notation. Second, we
introduce the concepts of Jones and Mueller matrices as classical
maps.
\subsection{Polarization states of light beams}
Many textbooks on classical optics introduce the concept of
polarized and unpolarized light  with the help of the Jones and
Stokes-Mueller calculi, respectively \cite{Damask}. In these
calculi, the description of classical polarization of light is
formally identical to the quantum description of pure and mixed
states of two-level systems, respectively \cite{Iso}.
In the Jones calculus, the electric field of a quasi-monochromatic
\emph{polarized} beam of light which propagates close the
$\bm{z}$-direction, is represented by a complex-valued
two-dimensional vector, the so-called \emph{Jones vector}
$\mathbf{E} \in \mathbb{C}^2: \mathbf{E} = E_0 \bm{x} + E_1
\bm{y}$, where the three real-valued unit vectors $\{
\bm{x},\bm{y},\bm{z}\}$ define an orthogonal Cartesian frame. The
same amount of information about the state of the field is also
contained in the $2 \times 2$ matrix $J$ of components $J_{ij} =
E_i E_j^*, \; (i,j=0,1)$, which is known as the \emph{coherency
matrix} of the beam \cite{BornWolf}. The matrix $J$ is Hermitean
and positive semidefinite
\begin{equation}\label{eq10}
J^\dagger = J, \qquad (\mathbf{v}, J \mathbf{v}) =
\left|(\mathbf{v}, \mathbf{E})\right|^2 \geq 0,
\end{equation}
where $\mathbf{v} \in \mathbb{C}^2$, and $(\mathbf{u}, \mathbf{v})
= \sum_{i=0}^{1} u_i^* v_i$ denotes the ordinary scalar product in
$\mathbb{C}^2$. Further, $J$ has the projection property
\begin{equation}\label{eq12}
J^2 = J \, \mathrm{Tr} J,
\end{equation}
 and its trace equals the total
intensity of the beam: $\mathrm{Tr} J = |E_0|^2 + |E_1|^2$. If we
choose the electric field units in such a way that $\tr J = 1$,
then $J$ has the same properties of a \emph{density matrix}
representing a two-level quantum system in a \emph{pure} state. In
classical polarization optics the coherency matrix description of
a light beam has the advantage, with respect to the Jones vector
representation, of generalizing to the concept of \emph{partially
polarized} light. Formally, the coherency matrix of a partially
polarized beam of light is characterized by the properties
(\ref{eq10}), while the projection property (\ref{eq12}) is lost.
In this case $J$ has the same properties of a density matrix
representing a two-level quantum system in a \emph{mixed} state.
Coherency matrices of partially polarized beams of light may be
obtained by tacking linear combinations $\sum_N w_N J_N$ of
coherency matrices $J_N$ of polarized beams (all parallel to the
same direction $\bm{z}$), where the index $N$ runs over an
\emph{ensemble} of field configurations and $w_N \geq 0$.
The degree of polarization (DOP, denoted $P$) of a partially
polarized beam is defined by the relation
\begin{equation}\label{eq13}
\mathrm{Det} J = (\tr J)^2(1 - P^2)/4.
\end{equation}
For a polarized beam of light, projection property (\ref{eq12})
implies $\mathrm{Det} J =0$ and $P = 1$, otherwise $0 \leq P < 1$.
 It should be noted that
the off-diagonal elements of the coherency matrix are
complex-valued and, therefore, not directly observables. However,
as any $2 \times 2$ matrix, $J$ can be written either in the Pauli
basis $X_\alpha$:
\begin{equation}\label{eq15}
\begin{array}{cclcccl}
X_0   &\equiv & \displaystyle{\left[\begin{array}{cc}
  1 & 0 \\
  0 & 1
\end{array}\right]}, &&
X_1   &\equiv &\displaystyle{ \left[\begin{array}{cc}
  0 & 1 \\
  1 & 0
\end{array}\right]},
\\\\
X_2   &\equiv &\displaystyle{ \left[\begin{array}{cc}
  0 & -i \\
  i & 0
\end{array}\right]}, &&
X_3   &\equiv &\displaystyle{ \left[\begin{array}{cc}
  1 & 0 \\
  0 & -1
\end{array}\right]},
\end{array}
%
  %  \qquad e15
\end{equation}
or in the {Standard} basis $Y_\alpha$:
\begin{equation}\label{eq17}
\begin{array}{cclcccl}
Y_0   &\equiv & \displaystyle{\left[\begin{array}{cc}
  1 & 0 \\
  0 & 0
\end{array}\right]}, &&
Y_1   &\equiv &\displaystyle{ \left[\begin{array}{cc}
  0 & 1 \\
  0 & 0
\end{array}\right]},
\\\\
Y_2   &\equiv &\displaystyle{ \left[\begin{array}{cc}
  0 & 0 \\
  1 & 0
\end{array}\right]}, &&
Y_3   &\equiv &\displaystyle{ \left[\begin{array}{cc}
  0 & 0 \\
  0 & 1
\end{array}\right]},
\end{array}
%
  %  \qquad e17
\end{equation}
as
\begin{equation}\label{eq20}
J = \frac{1}{2}\sum_{\alpha = 0}^3 x_\alpha X_\alpha = \sum_{\beta
= 0}^3 y_\beta Y_\beta,
\end{equation}
where $x_\alpha = \tr (X_\alpha J) \in \mathbb{R}$, $y_\beta = \tr
(Y_\beta^\dagger J) \in \mathbb{C}$ and, from now on, all Greek
indices $\alpha, \beta, \mu, \nu, \dots $, take the values
$0,1,2,3$. The four real coefficients $x_\alpha$, called the
\emph{Stokes parameters} \cite{NoteStokes} of the beam, can be
actually measured thus relating $J$ with observables of the
optical field. For example, $x_0 = \tr J$ represents the total
intensity of the beam. Conversely, the four complex coefficients
$y_\beta$ are not directly measurable but have the advantage to
furnish a particularly simple representation of the matrix $J$
since $y_0 = J_{00},  \, y_1 = J_{01}, \,  y_2 = J_{10}, \,  y_3 =
J_{11}$. The two different representations $x_\alpha$ and
$y_\beta$  are related via the  matrix
\begin{equation}\label{eq25}
V = \left[\begin{array}{cccc}
  1 & 0 & 0 & 1 \\
  0 & 1 & 1 & 0 \\
  0 & i & -i & 0 \\
  1 & 0 & 0 & -1
\end{array} \right] ,
\end{equation}
such that $x_\alpha = \sum_\beta V_{\alpha \beta} y_\beta$, where
$V_{\alpha \beta} = \tr(X_\alpha Y_\beta)$, and $V^\dagger V= 2
I_4= V V^\dagger $, where $I_4$ is the $4 \times 4$ identity
matrix.
\subsection{Polarization-transforming linear optical elements}
When a beam of light passes through an optical system its state of
polarization  may change. Within the context of polarization
optics, a polarization-affecting linear optical istrument is any
device that performs a \emph{linear} transformation upon the
electric field components of an incoming light beam without
affecting the spatial modes of the field. Half- and quarter-wave
plates,  phase shifters,  polarizers,  are all examples of such
devices. The class of polarization-affecting linear optical
elements comprises both non-depolarizing and depolarizing devices.
Roughly speaking, a \emph{non-depolarizing} linear optical element
transforms a polarized input beam into a polarized output beam. On
the contrary, a \emph{depolarizing} linear optical element
transforms a polarized input beam into a partially polarized
output beam \cite{LeRoy}.
A non-depolarizing device may be represented by a classical map
via a single $2 \times 2$ complex-valued matrix $T$,  the
\emph{Jones matrix} \cite{Damask}, such that
\begin{equation}\label{eq23}
\mathbf{E}_\mathrm{in} \rightarrow \mathbf{E}_\mathrm{out} = T
\mathbf{E}_\mathrm{in} ,
\end{equation}
for polarized input beams or, for light beams with arbitrary
degree of polarization:
\begin{equation}\label{eq22}
J_\mathrm{in} \rightarrow J_\mathrm{out} = T J_\mathrm{in}
T^\dagger.
\end{equation}
In this paper we consider only \emph{passive} (namely,
non-amplifying) optical devices for which the relation $\tr J_\out
\leq \tr J_\inp$ holds.
There exist two fundamental kinds of  non-depolarizing optical
elements,  namely \emph{retarders} and \emph{diattenuators}; any
other non-depolarizing element can be modelled as a retarder
followed by a diattenuator \cite{Lu96}. A retarder (also known as
\emph{birefringent} element) changes the \emph{phases} of the two
components of the electric-field vector of a beam, and may be
represented by a unitary Jones matrix $T_U$. A diattenuator (also
known as \emph{dichroic} element) instead changes the
\emph{amplitudes} of components of the electric-field vector
({polarization-dependent losses}), and may be represented by a
Hermitean Jones matrix $T_H$.

Let $\mathcal{T}_{\mathrm{ND}}$ denotes a generic non-depolarizing
device represented by  the Jones matrix $T$,  such that
$J_\mathrm{in} \rightarrow J_\mathrm{out} = T J_\mathrm{in}
T^\dagger$. We can rewrite explicitly this relation in terms of
components as
\begin{equation}\label{eq27}
(J_\out)_{ij}= T_{ik} T^*_{jl}(J_\inp)_{kl},
\end{equation}
where, from now on, summation over repeated indices is understood
and  all Latin indices $i,j,k,l,m,n, \ldots$ take the values $0$
and $1$. Since $ T_{ik} T^*_{jl} = (T\otimes T^*)_{ij,kl} \equiv
\mathcal{M}_{ij,kl} $ we can rewrite Eq. (\ref{eq27}) as
\begin{equation}\label{eq28}
(J_\out)_{ij}= \mathcal{M}_{ij,kl}(J_\inp)_{kl},
\end{equation}
where $\mathcal{M} = T\otimes T^*$ is a $4 \times 4$
complex-valued matrix  representing the device
$\mathcal{T}_{\mathrm{ND}}$, and the symbol $\otimes$ denotes the
ordinary Kronecker matrix product. $\mathcal{M}$ is also known as
the Mueller matrix in the Standard matrix basis \cite{AielloMath}
and it is simply related to the more commonly used real-valued
Mueller matrix $M$ \cite{Damask} via the change of basis matrix
$V$:
\begin{equation}\label{eq24}
 M = \frac{1}{2}V \mathcal{M} V^\dagger.
\end{equation}
For the present case of a non-depolarizing device, $M$ is named as
\emph{Mueller-Jones} matrix. From Eqs. (\ref{eq20}, \ref{eq28}) it
readily follows that we can indifferently represent the
transformation operated by $\mathcal{T}_{\mathrm{ND}}$ either in
the Standard or in the Pauli basis as
\begin{equation}\label{eq29}
y^\out_\alpha = \sum_{\beta = 0}^3 \mathcal{M}_{\alpha
\beta}y^\inp_\beta, \quad \mathrm{or} \quad x^\out_\alpha =
\sum_{\beta = 0}^3 M_{\alpha \beta} x^\inp_\beta,
\end{equation}
respectively.

With respect to the Jones matrix $T$, the Mueller matrices
$\mathcal{M}$ and $M$  have the advantage of generalizing to the
representation of \emph{depolarizing} optical elements.
Mueller matrices of depolarizing devices may be obtained by taking
linear combinations of Mueller-Jones matrices  of non-depolarizing
elements as
\begin{equation}\label{eq26}
\mathcal{M} = \sum_A p_A \mathcal{M}_A =  \sum_A p_A T_A \otimes
T_A^*,
\end{equation}
 where $p_A \geq 0$.
  Index $A$ runs over an ensemble (either deterministic \cite{Gil00} or stochastic
\cite{Kim87}) of  Mueller-Jones matrices $\mathcal{M}_A =  T_A
\otimes T_A^*$, each representing a non-depolarizing device. The
real-valued matrix $M$ corresponding to $\mathcal{M}$ written in
Eq. (\ref{eq26}), can be easily calculated by using Eq.
(\ref{eq24}) that it is still valid \cite{AielloMath}.
In the current literature $M$ is often written as \cite{Lu96}
\begin{equation}\label{eq30}
M = \left[%
\begin{array}{cc}
  M_{00} & \mathbf{d}^T \\
  \mathbf{p} & W \\
\end{array}%
\right],
\end{equation}
 where $\mathbf{p} \in \mathbb{R}^3, \; \mathbf{d} \in \mathbb{R}^3$, are known
as the \emph{polarizance vector} and the \emph{diattenuation
vector} (superscript $T$ indicates transposition), respectively,
and $W$ is a $3 \times 3$ real-valued matrix.
Note that
 $\mathbf{p}$ is zero for  pure depolarizers and pure
 retarders,
while $\mathbf{d}$ is nonzero only for dichroic optical elements
\cite{Lu96}.
Moreover, $W$ reduces to a three-dimensional orthogonal rotation
for pure retarders.
It the next Section, we shall show that  if we choose $M_{00}=1$
(this can be always done since it amounts to a trivial
\emph{polarization-independent} renormalization), the Mueller
matrix of a non-dichroic optical element ($\mathbf{d} =
\mathbf{0}$), is formally identical to a non-unital,
trace-preserving, one-qubit quantum map (also called channel)
\cite{Ruskai}. If also $\mathbf{p}=\mathbf{0}$ (pure depolarizers
and pure retarders), then $M$ is identical to a unital one-qubit
channel (as defined, e.g., in \cite{NielsenBook}).
\section{From classical to quantum maps: The spectral decomposition}
An important theorem in classical polarization optics states that
any linear optical element (either deterministic or stochastic) is
equivalent to a composite device made of at most four
non-depolarizing elements in parallel \cite{Anderson94}. This
theorem follows from the spectral decomposition of the Hermitean
positive semidefinite matrix $H$ \cite{Simon82} associated to
$\mathcal{M}$. In this Section we shortly review such theorem and
illustrate  its equivalence with the Kraus decomposition theorem
of one-qubit quantum maps \cite{NielsenBook}.

Given a Mueller matrix $\mathcal{M}$, it is possible to built a $4
\times 4 $ Hermitean positive semidefinite matrix $H =
H(\mathcal{M})$ by simply \emph{reshuffling} \cite{Zico} the
indices of $\mathcal{M}$:
\begin{equation}\label{eq32}
H_{ij,kl} \equiv \mathcal{M}_{ik,jl} = \sum_A p_A
(T_A)_{ij}(T_A^*)_{kl},
\end{equation}
where the last equality follows from Eq. (\ref{eq26}).
Equivalently, after introducing the composite indices $\alpha = 2
i +j,\, \beta = 2 k +l$,  we can rewrite Eq. (\ref{eq32}) as
$H_{\alpha \beta} = \sum_A p_A (T_A)_{\alpha}(T_A^*)_{\beta}$.
In view of the claimed connection between classical polarization
optics and one-qubit quantum mechanics, it worth noting that $H$
is formally identical to the \emph{dynamical} (or Choi) matrix,
describing a one-qubit quantum process \cite{Hdyna}.  The spectral
theorem for Hermitean matrices provides a \emph{canonical} (or
spectral) decomposition for $H$ of the form \cite{Horn1}
\begin{equation}\label{eq33}
H_{\alpha \beta} = \sum_{\mu = 0 }^3 \lambda_\mu
(\mathbf{u}_\mu)_{\alpha} (\mathbf{u}_\mu^*)_{\beta},
\end{equation}
 where $\lambda_\mu \geq 0$  are the
non-negative eigenvalues of $H$, and  $\{ \mathbf{u}_\mu \} = \{
\mathbf{u}_0, \, \mathbf{u}_1, \,\mathbf{u}_2, \,\mathbf{u}_3\}$
is the orthonormal basis of eigenvectors of $H$: $H \mathbf{u}_\mu
= \lambda_\mu \mathbf{u}_\mu$. Moreover, from a straightforward
calculation it follows that: $\sum_{\mu=0}^3 \lambda_\mu = 2
M_{00}$ \cite{AielloMath}.
 If we rearrange the four components
 of each eigenvector $\mathbf{u}_\mu$  to form a
 $2 \times 2$ matrices $T_\mu$ defined as
\begin{equation}\label{eq35}
T_\mu = \left[ \begin{array}{cc}
  (\mathbf{u}_\mu)_0 & (\mathbf{u}_\mu)_1 \\
  (\mathbf{u}_\mu)_2 & (\mathbf{u}_\mu)_3
\end{array} \right],
\end{equation}
we can rewrite Eq. (\ref{eq33}) as $H_{\alpha \beta} = \sum_\mu
\lambda_\mu (T_\mu)_{\alpha} (T_\mu^*)_{\beta}$. Since Eq.
(\ref{eq35}) can be rewritten as $(T_\mu)_{ij} =
(\mathbf{u}_\mu)_{\alpha
 =2i+j}$, we can go back from Greek to Latin indices and rewrite
 Eq. (\ref{eq33}) as
\begin{equation}\label{eq36}
H_{ij,kl} = \sum_{\mu =0}^3 \lambda_\mu (T_\mu)_{ij}
(T_\mu^*)_{kl} = \sum_{\mu =0}^3 \lambda_\mu (T_\mu \otimes
T_\mu^*)_{ik,jl}.
\end{equation}
 Finally,
from the  relation above and using Eq. (\ref{eq32}), we obtain
\begin{equation}\label{eq37}
\mathcal{M} = \sum_{\mu =0}^3 \lambda_\mu T_\mu \otimes T_\mu^*.
\end{equation}
Equation (\ref{eq37}) represents the content of the
\emph{decomposition theorem} in classical polarization optics, as
given by Cloude \cite{Anderson94,Cloude}. It implies, via Eq.
(\ref{eq28}), that the most general operation that a linear
optical device can perform upon a beam of light can be written as
\begin{equation}\label{eq40}
J_\inp \rightarrow J_\out = \sum_{\mu = 0}^3 \lambda_\mu T_\mu
J_\inp  T_\mu^\dagger,
\end{equation}
where the four Jones matrices $T_\mu$ represent four different
non-depolarizing optical elements.

Since $\lambda_\mu \geq 0$, Eq. (\ref{eq40}) is formally identical
to the Kraus form \cite{NielsenBook} of a completely positive
one-qubit quantum map $\mathcal{E}$. Therefore, because of the
isomorphism between $J$ and $\rho$ \cite{Iso}, when a single
photon encoding a polarization qubit (represented by the $2 \times
2$ density matrix $\rho_\inp$), passes through an optical device
classically described by the Mueller matrix $\mathcal{M} =
\sum_{\mu } \lambda_\mu T_\mu \otimes T_\mu^*$, its state will be
transformed according to
\begin{equation}\label{eq45}
 \rho_\inp \rightarrow \rho_\out \propto
  {\sum_{\mu = 0}^3 \lambda_\mu T_\mu \rho_\inp T_\mu^\dagger},
\end{equation}
where the proportionality symbol ``$\propto$'' accounts for a
possible renormalization to ensure $\tr \rho_\out = 1$. Such
renormalization is not necessary in the corresponding classical
equation (\ref{eq40}) since  $\tr J_\out$ is equal to the total
intensity of the output light beam that does not need to be
conserved.
Note that by using the definition  (\ref{eq37}) we can rewrite
explicitly  Eq. (\ref{eq45}) as
\begin{equation}\label{eq47}
\rho_{\out, ij} \propto \widetilde{\rho}_{\out, ij} =
\mathcal{M}_{ij,kl} \rho_{\inp, kl},
\end{equation}
where $(\rho)_{ij} =\langle i | \rho | j \rangle $ are density
matrix elements in the single-qubit standard basis $\{| i \rangle
\} $,  $i \in \{0,1 \} $, and $\widetilde{\rho}_\out$  is the
un-normalized single-qubit density matrix such that $\rho_\out =
\widetilde{\rho}_\out / \tr \widetilde{\rho}_\out$.
From Eqs. (\ref{eq24}-\ref{eq30}) and Eq. (\ref{eq47}), it readily
follows
\begin{eqnarray}
\tr  \widetilde{\rho}_\out  &  = &  M_{00}
+ M_{01}(\rho_{\inp, 01} + \rho_{\inp, 10}) \nonumber\\
&  &+  i M_{02} (\rho_{\inp, 01} - \rho_{\inp, 10})\nonumber\\
&  & + M_{03}(\rho_{\inp, 00} - \rho_{\inp, 11}), \label{eq48}
\end{eqnarray}
where we have assumed $\tr \rho_{\inp}=1 $. The equation above
shows that $\mathcal{M}$ represents a trace-preserving map only if
$M_{00}=1$ and $\mathbf{d}^T = (M_{01},M_{02},M_{03}) = (0,0,0)$,
namely, only if $\mathcal{M}$ describes the action of a
non-dichroic optical instrument. In addition, if $\rho_{\inp}$
represents a completely mixed state, that is if $\rho_{\inp} =
X_0/2$, then from  Eq. (\ref{eq47}) it follows:
\begin{eqnarray}\label{eq49}
\widetilde{\rho}_\out = \frac{1}{2} \sum_{\mu = 0}^3 p_\mu X_\mu,
\end{eqnarray}
were we have defined $p_0 \equiv M_{00}$ and $( p_1,p_2,p_3) =
\mathbf{p}$ is the polarizance vector. Equation (\ref{eq49}) shows
that in this case $\tr \widetilde{\rho}_\out =  M_{00}$, and
$\rho_\out = \widetilde{\rho}_\out /  M_{00} \neq X_0/2$ if
$\mathbf{p} \neq \mathbf{0}$, that is, the map represented by
$\mathcal{M}$ (or, $M$) is unital only if $\mathbf{p}=0$.

By writing Eqs. (\ref{eq40}-\ref{eq49}) we have thus completed the
review of the analogies between linear optics and one-qubit
quantum maps. In the next Section we shall study the connection
between classical polarization optics and \emph{two-qubit} quantum
maps.
\section{Polarization optics and two-qubit quantum maps}
Let us consider a typical SPDC setup where pairs of photons are
created in the quantum state $\rho$ along two well defined spatial
modes (say, path $A$ and path $B$) of the electromagnetic field,
as shown in Fig. 1.
%
%%%%%%%%%%%%%%%%%%%%%%%%%%%%%%%%%%%%%%%%%%%%%%%%%%%%%%%%%%%%%%%%%%%%%%%%%%%%%%%%%%%%
\begin{figure}[!ht]
\includegraphics[angle=0,width=7truecm]{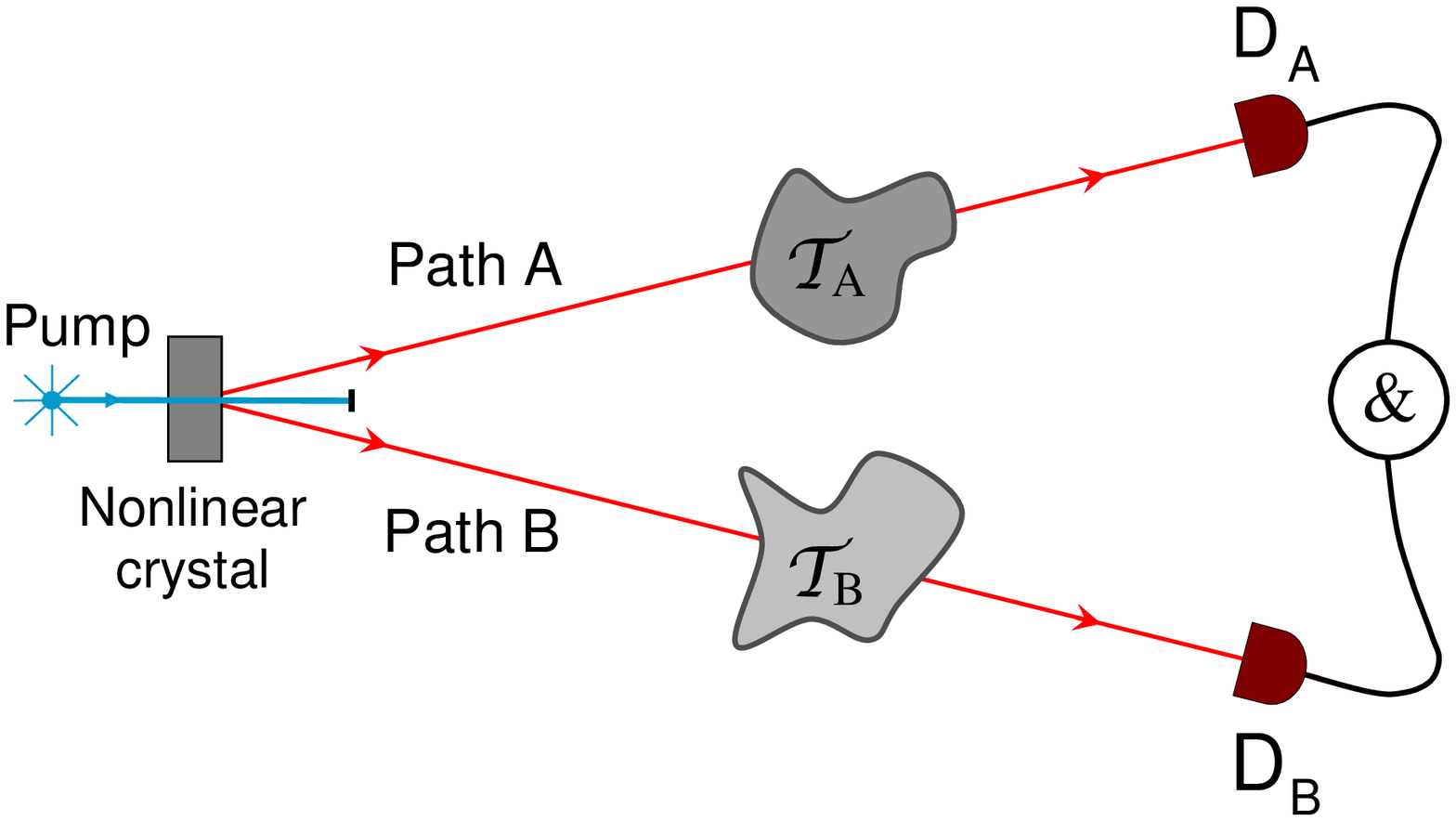}
\caption{\label{fig:1}(Color online) Layout of a typical SPDC
experimental setup. An optically pumped \textsf{ nonlinear
crystal}, emits photon pairs that propagate along path \textsf{A}
and \textsf{B} through  the scattering devices $\mathcal{T}_A$ and
$\mathcal{T}_B$, respectively. Scattered photons are detected in
coincidence by detectors $\mathrm{\textsf{D}_\textsf{A}}$ and
$\mathrm{\textsf{D}}_\mathrm{\textsf{B}}$ that permit a
tomographically complete  two-photon polarization state
reconstruction.}
\end{figure}
%%%%%%%%%%%%%%%%%%%%%%%%%%%%%%%%%%%%%%%%%%%%%%%%%%%%%%%%%%%%%%%%%%%%%%%%%%%%%%%%%%%%%
%
Each photon of the pair encodes a polarization qubit and $\rho$
can be represented by a $4 \times 4$ Hermitean matrix. Let
$\mathcal{T}_A$ and $\mathcal{T}_B$ be two \emph{distinct} optical
devices put across path $A$ and path $B$, respectively. Their
action upon the two-qubit state  $\rho$ can be described by a
\emph{bi-local} quantum map $\rho \rightarrow \mathcal{E}_A
\otimes \mathcal{E}_B [\rho] $ \cite{Ziman1}.
A sub-class of bi-local quantum maps occurs when either
$\mathcal{T}_A$ or $\mathcal{T}_B$ is \emph{not} present in the
setup, then either $\mathcal{E}_A = \mathcal{I}$ or $\mathcal{E}_B
= \mathcal{I}$, respectively, and the corresponding map is said to
be \emph{local}. In the above expressions $ \mathcal{I}$
represents the identity map: It does not change \emph{any} input
state.
 When a map is local, that is when it acts on a
single qubit, it is subjected to some restrictions. This can be
easily understood in the following way: For definiteness, let
assume $\mathcal{E}_B = \mathcal{I}$ so that the local map
$\mathcal{E}$ can be written as $\mathcal{E} [\rho]=\mathcal{E}_A
\otimes \mathcal{I} [\rho]$. Let Alice and Bob be two spatially
separated observer who can detect qubits in  modes $A$ and $B$,
respectively, and let $\rho$ and $\rho_\mathcal{E}$ denote
 the two-qubit quantum state before and after $\mathcal{T}_A$, respectively.
 In absence of any causal
connection between photons in path $A$ with photons in path $B$,
special relativity demands that Bob cannot detect via any type of
local measurement the presence of the device $\mathcal{T}_A$
located in path  $A$. Since the state of each qubit received by
Bob is represented by the reduced density matrix
$\rho_\mathcal{E}^B = \mathrm{Tr}|_A (\rho_\mathcal{E})$, the
locality constraint can be written as
\begin{equation}\label{i10}
\rho_\mathcal{E}^B = \rho^B.
\end{equation}
We can write explicitly the map $\mathcal{E}_A \otimes
\mathcal{I}$ as a Kraus operator-sum decomposition
\cite{NielsenBook}
\begin{equation}\label{i20}
\rho \mapsto \rho_\mathcal{E} \propto \sum_{\mu=0}^3 \lambda_\mu
\left(A_\mu \otimes I \right) \rho \left(A_\mu^\dagger \otimes I
\right),
\end{equation}
where,  from now on, the symbol $I$ denotes the $2 \times2$
identity matrix and  $\{ A_\mu \}$ is a set of four $2 \times 2$
Jones matrices describing the action of  $\mathcal{T}_A$. Then,
Eq. (\ref{i10}) becomes
\begin{equation}\label{i30}
\sum_{k,l}\rho_{li,kj} \Bigl( \sum_{\mu=0}^3 \lambda_\mu
A_\mu^\dagger A_\mu \Bigr)_{kl} \propto \sum_{k} \rho_{ki,kj},
\end{equation}
which implies the \emph{trace-preserving} condition on the local
map $\mathcal{E}_A \otimes \mathcal{I}$:
\begin{equation}\label{i40}
\sum_{\mu=0}^3 \lambda_\mu A_\mu^\dagger A_\mu \propto I.
\end{equation}
 Local maps that do not satisfy Eq.
(\ref{i40}) are classified as \emph{non-physical}. In this Section
we show how to associate a general two-qubit quantum map
$\mathcal{E} [\rho]=\mathcal{E}_A \otimes \mathcal{E}_B [\rho]$ to
the classical Mueller matrices $\mathcal{M}^A$ and $\mathcal{M}^B$
 describing the optical devices $\mathcal{T}_A$ and
$\mathcal{T}_B$, respectively. Surprisingly, we shall find that do
exist \emph{physical} linear optical devices (dichroic elements)
that may generate \emph{non-physical} two-qubit quantum maps
\cite{Aiello062}.

Let denotes with $| ij\rangle \equiv |i \rangle \otimes |j
\rangle, \; i,j \in \{0,1\}$ the two-qubit standard basis.
A pair of qubits is initially prepared in the generic state $ \rho
= \rho_{ij,kl} |ij\rangle \langle kl| = \rho_{ik,jl}^R |i\rangle
\langle k| \otimes |j\rangle \langle l| $, where superscript $R$
indicates {reshuffling}  of the indices, the same operation we
used to pass from $\mathcal{M}$ to $H$: $\rho_{ik,jl}^R \equiv
\rho_{ij,kl} = \langle i j | \rho | k l \rangle$. $\rho$ is
transformed under the action of the \emph{bi-local} linear map
$\mathcal{E}[\rho]=\mathcal{E}_A \otimes \mathcal{E}_B[\rho]$ into
the state
\begin{equation}\label{eq50}
\rho_\mathcal{E} = \mathcal{E}_A \otimes \mathcal{E}_B[\rho]
\propto \sum_{\mu,\nu} \lambda_\mu \lambda_\nu  \bigl(  A_\mu
\otimes B_\nu \bigr) \, \rho \, \bigl( A_\mu^\dagger \otimes
B_\nu^\dagger \bigr),
\end{equation}
where $\{A_\mu\}$ and $\{B_\nu\}$ are two sets of $2 \times 2$
Jones matrices describing the action of $\mathcal{T}_A$ and
$\mathcal{T}_B$, respectively.  From Eq. (\ref{eq50}) we can
calculate explicitly the matrix elements $\langle i j |
\rho_\mathcal{E} | k l \rangle = (\rho_\mathcal{E})_{ij,kl} $ in
the two-qubit standard basis:
%
% By writing explicitly Eq. (\ref{eq50}) in the two-qubit basis
%$| ij\rangle $, it readily follows
%
\begin{equation}\label{eq55}
\begin{array}{rcl}
  (\rho_\mathcal{E})_{ij,kl} & \propto &

  \lambda_\mu (A_\mu)_{im} (A_\mu^*)_{kp} \,\rho^R_{mp,nq} \,
 \lambda_\nu (B_\nu)_{jn} (B_\nu^*)_{lq}\\\\
   & = & \mathcal{M}^A_{ik,mp} \mathcal{M}^B_{jl,nq}
   \rho^R_{mp,nq},
\end{array}
\end{equation}
 where
summation over repeated Latin and Greek indices is understood.
Since by definition $(\rho_\mathcal{E})_{ij,kl} =
(\rho_\mathcal{E}^R)_{ik,jl}$ we can rewrite Eq. (\ref{eq55})
using only Greek indices as
\begin{equation}\label{eq57}
(\rho_\mathcal{E}^R)_{\alpha \beta} \propto \mathcal{M}^A_{\alpha
\mu} \mathcal{M}^B_{\beta \nu} \, \rho^R_{\mu \nu} = \left(
\mathcal{M}^A \otimes \mathcal{M}^B \right)_{\alpha \beta, \mu
\nu} \rho^R_{\mu \nu},
\end{equation}
 where
summation over repeated  Greek indices is again understood.
Equation (\ref{eq57}) relates classical quantities (the two
Mueller matrices $\mathcal{M}^A$ and $\mathcal{M}^B$) with quantum
ones (the input and output density matrices $\rho^R$ and
$\rho_\mathcal{E}^R$, respectively). Moreover, it is easy to see
that  Eq. (\ref{eq57}) is the two-qubit quantum analogue of  Eq.
(\ref{eq29}).
In fact, if we introduce the $16 \times 16$ two-qubit Mueller
matrix $\mathcal{M} \equiv \mathcal{M}^A \otimes \mathcal{M}^B$,
and the input and output two-qubit Stokes parameters in the
standard basis defined as: $ y^\inp_{b = 4 \mu +\nu}=
(\rho^R)_{\mu \nu}$, $y^\out_{a = 4 \alpha +\beta}=
(\rho_\mathcal{E}^R)_{\alpha \beta}$, where $a,b \in \{0, \ldots,
15\}$, then we can write Eq. (\ref{eq57})  as
\begin{equation}\label{eq59}
y^\out_a \propto \sum_{b = 0}^{15} \mathcal{M}_{ab} y^\inp_b,
\end{equation}
which is formally identical to
 Eq. (\ref{eq29}). Thus,  Eq. (\ref{eq59}) realizes
 the connection between classical polarization optics and
 two-qubit quantum maps.

An important case occurs when $\mathcal{E}_{B} =
\mathcal{I}\Rightarrow $ $\mathcal{M}^{B} = I_4$ and Eq.
(\ref{eq57}) reduces to
\begin{equation}\label{eq54}
\rho_\mathcal{E}^R  \propto \mathcal{M}^A \rho^R.
\end{equation}
Equation (\ref{eq54})  illustrates once more the simple relation
existing between the \emph{classical} Mueller matrix
$\mathcal{M}^A$ and the \emph{quantum} state $\rho_\mathcal{E}$.

With a typical SPDC setup it is not difficult to prepare pairs of
entangled photons in the singlet polarization state. Via a direct
calculation, it is simple to show that when $\rho$ represents two
qubits in the singlet state $\rho_s = \frac{1}{4}(X_0 \otimes X_0
- X_1 \otimes X_1 - X_2 \otimes X_2 - X_3 \otimes X_3)$ and
$\mathcal{M}^A$ is normalized in such a way that $M^A_{00}=1$,
then the proportionably symbol in the last equation above can be
substituted with the equality symbol:
\begin{equation}\label{eq60}
\rho_\mathcal{E}^R =  \mathcal{M} \rho^R_s \quad \Longrightarrow
\quad  \rho_\mathcal{E} =  \left( \mathcal{M} \rho^R_s \right)^R,
\end{equation}
where, from now on,  we write $\mathcal{M}$ for $\mathcal{M}^A$ to
simplify the notation.
 Note that this
pleasant property is true not only or the singlet but for all four
Bell states \cite{NielsenBook}, as well. Equation (\ref{eq60}) has
several remarkable consequences: Let $M$ denotes the real-valued
Mueller matrix associated to $\mathcal{M}$ and assume $M_{00}=1$.
Then, the following results hold:
\begin{eqnarray}
 \tr({\rho}_\mathcal{E}^2)   & =  &\tr(M {M}^T)/4, \label{eq80} \\
\tr|_A (\widetilde{\rho}_\mathcal{E})  &  = &(A + D) + M_{01}(B+C) \nonumber\\
&  &+  i M_{02} (B - C) + M_{03}(A - D), \label{eq90}
\end{eqnarray}
where $\widetilde{\rho}_\mathcal{E} \equiv \left( \mathcal{M}
\rho^R \right)^R$ is the un-normalized output density matrix.
Equation (\ref{eq90}) is more general than Eq. (\ref{eq80}), since
it holds for \emph{any} input density matrix $\rho$ and not only
for the singlet one $\rho_s$. In addition, in Eq. (\ref{eq90})  we
wrote the input density matrix $\rho$ in a block-matrix form as
\begin{equation}\label{eq100}
\rho =  \left[ \begin{array}{cc}
  A & B \\
  C & D
\end{array}  \right],
\end{equation}
where $A$, $B$, $C= B^\dagger$, and $D$ are $2 \times 2$
sub-matrices and $A + D= \tr|_A (\rho)$.
Equation (\ref{eq80}) shows that the degree of mixedness of the
quantum state $\rho_\mathcal{E}$ is in a one-to-one correspondence
with the classical depolarizing power \cite{LeRoy} of the device
represented by $M$. Finally, Eq. (\ref{eq90}), together with Eqs.
(\ref{eq30},\ref{i10}), tells us that the two-qubit quantum map
Eq. (\ref{eq60}) is trace-preserving only if the device is not
dichroic, namely only if $\mathbf{d}^T = (M_{01},M_{02},M_{03})=
(0,0,0)$. This last result shows that despite of their physical
nature (think of, e.g., a polarizer), dichroic optical elements
 must be handled with care when used to build two-qubit
quantum maps. We shall discuss further this point in the next
Section.

Before concluding this Section, we want to point out the analogy
between the $16 \times 16$ Mueller matrix $\mathcal{M} =
\mathcal{M}^A \otimes \mathcal{M}^B$ associated to a bi-local
\emph{two}-qubit quantum map, and the $4 \times 4$ Mueller-Jones
matrix $\mathcal{M} = T \otimes T^*$ representing a
non-depolarizing device in a \emph{one}-qubit quantum map. In both
cases the Mueller matrix is said to be \emph{separable}. Then, in
Eq. (\ref{eq26}) we learned how to build non-separable Mueller
matrices representing depolarizing optical elements. By analogy,
we can now build non-separable two-qubit Mueller matrices
representing \emph{non-local} quantum maps, as
\begin{equation}\label{eq102}
\mathcal{M} = \sum_{A,B} w_{AB}\mathcal{M}^A \otimes
\mathcal{M}^B,
\end{equation}
where $w_{AB} \geq 0$, $w_{AB} \neq w_A \times w_B$, and indices
$A,B$ run over two ensembles of arbitrary Mueller matrices
$\mathcal{M}^A$ and $\mathcal{M}^B$ representing optical devices
located in path $A$ and path $B$, respectively.
\section{Applications}
In this Section we exploit our formalism, by applying it to two
different cases. As a first application, we build a  simple
phenomenological model capable to explain certain of our recent
experimental results \cite{Puentes06b} about scattering of
entangled photons. The second application consists in the explicit
construction of a bi-local quantum map generating two-qubit MEMS
states. A realistic physical implementation of such map is also
given.
\subsection{Example 1: A simple phenomenological model}
In Ref. \cite{Puentes06b}, by using a setup similar to the one
shown in Fig. 1, we have experimentally generated entangled
two-qubit mixed states that lie upon and below the Werner curve in
the linear entropy-tangle plane \cite{Peters}.
In particular, we have found that: {(a)} Birefringent scatterers
 always produce \emph{generalized Werner
states} of the form $\rho_{GW}=V\otimes I \rho_{W}
V^{\dagger}\otimes I$, where $\rho_{W}$ denotes ordinary Werner
states  \cite{Werner}, and $V$ represents an arbitrary unitary
operation; (b) Dichroic scatterers generate sub-Werner states,
that is states that lie below the Werner curve in the linear
entropy-tangle plane.
In both cases, the input photon pairs were experimentally prepared
in the polarization singlet state $\rho_s$.
 In this
subsection we build, with the aid of Eq. (\ref{eq60}), a
 phenomenological model explaining both results (a) and (b).

 To this end let us consider the experimental
setup represented in Fig. 1.   According to the actual scheme used
in Ref. \cite{Puentes06b}, where a single scattering device was
present, in this Subsection we assume $\mathcal{T}_B =
\mathcal{I}$, so that the resulting quantum map is local. The
scattering  element $\mathcal{T}_A$ inserted across path $A$ can
be classically described by some Mueller matrix $\mathcal{M}$. In
Ref. \cite{Lu96}, Lu and Chipman have shown that \emph{any} given
Mueller matrix $\mathcal{M}$ can be decomposed in the product
\begin{equation}\label{eq110}
\mathcal{M} = \mathcal{M}_D \mathcal{M}_B  \mathcal{M}_\Delta,
\end{equation}
where $\mathcal{M}_\Delta$,  $\mathcal{M}_B$, and  $\mathcal{M}_D$
are complex-valued Mueller matrices representing a pure
depolarizer, a retarder, and a diattenuator, respectively. Such
decomposition is not unique, for example, $\mathcal{M} =
  \mathcal{M}_\Delta \mathcal{M}_D \mathcal{M}_B$ is another valid decomposition \cite{Morio04}.
   Of course, the actual values of
$\mathcal{M}_\Delta$, $\mathcal{M}_B$, and $\mathcal{M}_D$ depend
on the specific order one chooses. However, in any case they have
the general forms given below:
\begin{eqnarray}
\mathcal{M}_\Delta & = & \left[ \begin{array}{cccc}
  \frac{1+c}{2} & 0 & 0 & \frac{1-c}{2} \\
  0 &  \frac{a+b}{2} & \frac{a-b}{2} & 0 \\
  0 & \frac{a-b}{2} & \frac{a+b}{2} & 0 \\
   \frac{1-c}{2} & 0 & 0 & \frac{1+c}{2}
\end{array} \right],
\label{eq120}\\
\mathcal{M}_B & = & T_U \otimes T_U^*  , \label{eq130}\\
 \mathcal{M}_D & = & T_H \otimes T_H^* ,\label{eq140}
\end{eqnarray}
 where $a,b,c \in \mathbb{R}$, and $T_U$,
$T_H$ are the unitary and Hermitean  Jones matrices representing a
retarder and a diattenuator, respectively.
Actually, the expression of $\mathcal{M}_\Delta$ given in Eq.
(\ref{eq120}) is not the most general possible \cite{Lu96}, but it
is the correct one for the representation of pure depolarizers
with zero polarizance, such as the ones used in  Ref.
\cite{Puentes06b}.
Note that although $\mathcal{M}_B$ and $\mathcal{M}_D$ are
Mueller-Jones matrices, $\mathcal{M}_\Delta$ is not. When $a=b=c
\equiv p: \, p \in [0,1] $ the depolarizer is said to be
\emph{isotropic} (or, better, polarization-isotropic). This
 case is particularly relevant when birefringence and dichroism are absent. In this
 case $\mathcal{M}_B = I_4 = \mathcal{M}_D$, and Eq. (\ref{eq110})
 gives $\mathcal{M} = \mathcal{M}_\Delta$. Thus, by using Eq.
(\ref{eq120}) we can calculate $\mathcal{M}_\Delta (p)$ and use it
in Eq. (\ref{eq60}) to obtain
\begin{equation}\label{eq150}
\rho_\mathcal{E} = p \rho_s + \frac{1 - p}{4} I_4 \equiv \rho_W,
\end{equation}
that is, we have just obtained  a Werner state: $\rho_\mathcal{E}
= \rho_W$! Thus, we have found that  a local
polarization-isotropic scatterer acting upon the two-qubit singlet
state, generates Werner states.

Next, let us consider the cases of birefringent (retarders) and
dichroic (diattenuators) scattering devices that we used in our
experiments. In these cases the total Mueller matrices
$\mathcal{M}$ of the devices under consideration, can be written
as $\mathcal{M} = \mathcal{M}_Z \mathcal{M}_\Delta$, where either
$Z=B$ or $Z = D$, and $\mathcal{M}_\Delta= \mathcal{M}_\Delta(p)$
represents a polarization-isotropic depolarizer. For definiteness,
let consider in detail only the case of a birefringent scatterer,
since the case of a dichroic one can be treated in the same way.
In this case
\begin{equation}\label{eq160}
\mathcal{M}_B \mathcal{M}_\Delta(p) = \sum_{\mu = 0}^3
\lambda_\mu(p) T_U T_\mu \otimes T_U ^* T_\mu^*,
\end{equation}
and, as result of a straightforward calculation, $\lambda_0 = (1 +
3p)/2, \; \lambda_1=\lambda_2=\lambda_3=(1-p)/2$, $T_\mu = X_\mu
/\sqrt{2}$; while $T_U $ is an arbitrary unitary $2 \times 2$
Jones matrix representing a generic retarder. For the sake of
clarity, instead of using directly Eq. (\ref{eq60}), we prefer to
rewrite Eq. (\ref{eq50}) adapted to this case as
\begin{eqnarray}\label{eq170}
\rho_\mathcal{E} & = & \sum_{\mu = 0}^3 \lambda_\mu (p)  \bigl(
T_U T_\mu\otimes I \bigr) \, \rho_s \, \bigl(   T_\mu^\dagger T_U
^\dagger
\otimes I \bigr) \nonumber \\
&  = & T_U  \otimes I \left[ \sum_{\mu = 0}^3 \lambda_\mu (p)
\bigl( T_\mu\otimes I \bigr) \, \rho_s \, \bigl(  T_\mu^\dagger
\otimes I
\bigr)\right]T_U ^\dagger \otimes I \nonumber \\
&  = & T_U  \otimes I \rho_W T_U ^\dagger \otimes I \nonumber \\
&  = & \rho_{GW},
\end{eqnarray}
where Eq. (\ref{eq150}) has been used. Equation Eq. (\ref{eq170})
clearly shows that the effect of a birefringent scatterer is to
generate what we called \emph{generalized Werner states}, in full
agreement with our  experimental results \cite{Puentes06b}.

The analysis for the case of a dichroic scatterer can be done in
the same manner leading to the result
\begin{eqnarray}\label{eq180}
\rho_\mathcal{E} \propto \widetilde{\rho}_\mathcal{E}  = T_H
\otimes I \rho_W T_H ^\dagger \otimes I,
\end{eqnarray}
where $T_H$ is a $2 \times 2$ Hermitean matrix representing a
generic diattenuator \cite{Damask}:
\begin{eqnarray}\label{eq190}
T_H = \left[ \begin{array}{cc}
  d_0 \cos\theta^2 +
              d_1\sin\theta^2 & (d_0 - d_1)\cos\theta\sin\theta \\
(d_0 - d_1)\cos\theta\sin\theta & d_1\cos\theta\ ^2 + d_0
\sin\theta^2
\end{array} \right],
\end{eqnarray}
where $d_i \in [0,1]$, are the diattenuation factors, while
$\theta \in (0, 2 \pi ]$ gives the direction of the transmission
axis of the linear polarizer to which $T_H$ reduces when either
$d_0=0$ or $d_1 = 0$. Figure 2 reports, in the tangle-linear
entropy plane, the results of a numerical simulation were we
generated $10^4$ states $\rho_\mathcal{E}$ from Eq. (\ref{eq180}),
by randomly generating (with uniform distributions) the four
parameters $p, \, d_0,\, d_1$, and $\theta$ in the ranges: $p \,
,d_0\, ,d_1  \in [0,1]$, $\theta \in (0, 2 \pi ]$.
%
%%%%%%%%%%%%%%%%%%%%%%%%%%%%%%%%%%%%%%%%%%%%%%%%%%%%%%%%%%%%%%%%%%%%%%%%%%%%%%%%%%%%%%%%%%%
\begin{figure}[ht!]
\includegraphics[angle=0,width=6truecm]{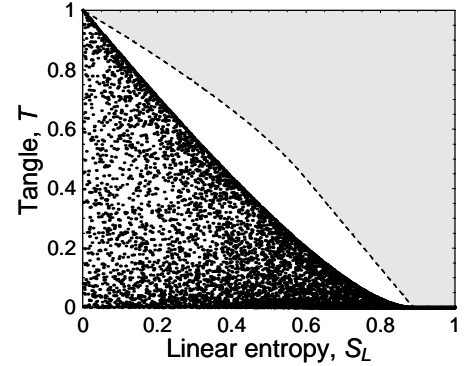}
\caption{\label{fig:2} Numerical simulation from our
phenomenological model qualitatively reproducing the behavior of a
dichroic scattering system. The gray region represents unphysical
states and it is bounded from below by MEMS (dashed curve). The
lower continuous thick curve represents Werner states.  }
\end{figure}
%%%%%%%%%%%%%%%%%%%%%%%%%%%%%%%%%%%%%%%%%%%%%%%%%%%%%%%%%%%%%%%%%%%%%%%%%%%%%%%%%%%%%%%%%%%%%
%
The numerical simulation shows that a local dichroic scatterer may
generate \emph{sub-Werner} two-qubit states, that  is states
located \emph{below} the Werner curve in the tangle-linear entropy
plane.  The qualitative agreement between the result of this
simulation and the experimental findings shown in Fig. 3  of Ref.
\cite{Puentes06b} is evident.
\subsubsection{Discussion}
It should be noticed that while we used the equality symbol in
writing Eq.  (\ref{eq170}), we had to use  the proportionality
symbol in writing Eq. (\ref{eq180}). This is a consequence of the
Hermitean character of the Jones matrix $T_H$ that generates a
\emph{non-trace-preserving} map. In fact, in this case from $M =
M_D M_\Delta(p)$, where $M_D = (V T_H \otimes T_H^* V^\dagger)/2$
and $M_\Delta(p) = [V \mathcal{M}_\Delta(p) V^\dagger]/2$ [see Eq.
(\ref{eq24})], we obtain
$\mathrm{Tr}(\widetilde{\rho}_\mathcal{E}) = (d_0^2 + d_1^2)/2
\neq 1$. Moreover,  Eq.  (\ref{eq90}) gives
\begin{equation}\label{eq191}
\begin{array}{rcl}
{\rho}_\mathcal{E}^B & = & \displaystyle{\mathrm{Tr}|_A({\rho}_\mathcal{E})}\\\\
& = &\displaystyle{  \frac{X_0}{2} - p \left(  \frac{d_0^2
-d_1^2}{d_0^2 + d_1^2} \right)
 \frac{X_1  \sin 2\theta + X_3\cos 2 \theta}{2}},
\end{array}
\end{equation}
where ${\rho}_\mathcal{E} =
\widetilde{\rho}_\mathcal{E}/\mathrm{Tr}(\widetilde{\rho}_\mathcal{E})$.
This result is in contradiction, for  $d_0 \neq d_1$, with the
locality constraint expressed by Eq. (\ref{i10}) which requires
\begin{equation}\label{eq192}
\rho_\mathcal{E}^B =  \frac{X_0}{2}.
\end{equation}
As we already discussed in the previous Section, only the latter
result seems to be physically meaningful since photons in path
$B$, described by $\rho_\mathcal{E}^B$, cannot carry information
about device $\mathcal{T}_A$ which is located across path $A$. On
the contrary, Eq. (\ref{eq191}) shows that $\rho_\mathcal{E}^B$ is
expressed in terms of the four physical parameters $p, d_0, d_1$
and $\theta$ that characterize $\mathcal{T}_A$. Is there a
contradiction here?

In fact, there is none! One should keep in mind that Eq.
(\ref{eq191}) expresses  the \emph{one-qubit} reduced density
matrix $\rho_\mathcal{E}^B$ that is extracted
%[via  Eq. (\ref{eq90})]
from the \emph{two-qubit} density matrix $\rho_\mathcal{E}$
\emph{after} the latter has been reconstructed by the two
observers Alice and Bob  by means of \emph{nonlocal} coincidence
measurements. Such matrix contains information about \emph{both}
qubits and, therefore, contains also information about
$\mathcal{T}_A$.
 Conversely, $\rho^B_\mathcal{E}= X_0/2$ in  Eq.
(\ref{eq192}), is the reduced density matrix that could be
reconstructed by Bob alone via \emph{local} measurements
\emph{before} he and Alice had compared their own experimental
results and had selected from the raw data the coincidence counts.

From a physical point of view, the discrepancy between Eq.
(\ref{eq191}) and Eq. (\ref{eq192}) is due to the
\emph{polarization-dependent} losses (that is, $d_0 \neq d_1$)
that characterize dichroic optical devices and it is unavoidable
when such elements are present in an experimental setup. Actually,
it has been already noticed that
 a dichroic optical element  necessarily performs a kind of post-selective
 measurement \cite{Brunner03}. In our case
 coincidence measurements post-select only those
photons that have not been absorbed by the dichroic elements
present in the setup.
However, since in any SPDC setup even the initial singlet state
 is actually a post-selected state (in order to cut off the
 otherwise
overwhelming vacuum contribution), the practical use of dichroic
devices does not represent a severe limitation for such setups.
\subsection{Example 2: Generation of two-qubit MEMS states}
In the previous subsection we have shown that it is possible to
generate  two-qubit states represented by points upon and below
the Werner curve in the tangle-linear entropy plane, by operating
on a single qubit (local operations) belonging to a pair initially
prepared in the entangled singlet state. In another paper
\cite{Aiello062} we have shown that it is also possible to
generate MEMS states (see, e.g., \cite{Peters,Barbieri} and
references therein), via local operations. However, the price to
pay in that case was the necessity to use a dichroic device that
could not be represented by a ``physical'', namely a
trace-preserving, quantum map.
In the present subsection, as an example illustrating the
usefulness of our conceptual scheme, we show that by allowing
bi-local operations performed by two separate optical devices
$\mathcal{T}_A$ and $\mathcal{T}_B$ located as  in Fig. 1, it is
possible to achieve MEMS states \emph{without} using dichroic
devices.

To this end, let us start by rewriting explicitly Eq.
(\ref{eq50}), where the most general bi-local quantum map $
\mathcal{E}[\rho] = \mathcal{E}_A \otimes \mathcal{E}_B[\rho]$
operating upon the generic input two-qubit state $\rho$, is
represented by a Kraus decomposition:
\begin{equation}\label{eq210}
\rho_\mathcal{E} = \mathcal{E}_A \otimes \mathcal{E}_B[\rho] =
\sum_{\mu,\nu} \lambda_\mu \lambda_\nu  \bigl(  A_\mu \otimes
B_\nu \bigr) \, \rho \, \bigl( A_\mu^\dagger \otimes B_\nu^\dagger
\bigr),
\end{equation}
where now the equality symbol can be used since we assume that
both single-qubit maps $\mathcal{E}_A$ and $\mathcal{E}_B$ are
 trace-preserving,
\begin{equation}\label{eq220}
\sum_{\mu=0}^3  \lambda_\mu A_\mu^\dagger A_\mu = I  =
\sum_{\nu=0}^3 \lambda_\nu B_\nu^\dagger B_\nu,
\end{equation}
but not necessarily unital: $\mathcal{E}_F[I]\neq I, \, F \in
\{A,B\} $ \cite{Ziman1}. Under the action of $\mathcal{E}$, the
initial state of each  qubit travelling in path $A$ or path $B$ is
transformed into either the output state
\begin{eqnarray}
\rho_\mathcal{E}^A  = \mathrm{Tr}|_B (\rho_\mathcal{E}) =
\sum_{\mu=0}^3   \lambda_\mu A_\mu \rho^A A_\mu^\dagger ,
\label{eq230}
\end{eqnarray}
or
\begin{eqnarray}
\rho_\mathcal{E}^B  = \mathrm{Tr}|_A (\rho_\mathcal{E}) =
\sum_{\nu=0}^3  \lambda_\nu B_\nu \rho^B B_\nu^\dagger,
\label{eq240}
\end{eqnarray}
respectively, where $\rho^A =  \mathrm{Tr}|_B (\rho)$, and
$\rho^B = \mathrm{Tr}|_A (\rho)$.
Without loss of generality, we assume that the two qubits are
initially prepared in the singlet state:  $\rho = \rho_s$. Then
Eqs. (\ref{eq230}-\ref{eq240}) reduce to $\rho_\mathcal{E}^F =
\sum_\alpha   F_\alpha  F_\alpha^\dagger/2, \; F \in \{A,B\}$.
From the previous analysis [see Eqs. (\ref{eq50}-\ref{eq57})] we
know that to each bi-local \emph{quantum} map $\mathcal{E}_A
\otimes \mathcal{E}_B$ can be associated a pair of
\emph{classical} Mueller  matrices $\mathcal{M}_A$ and
$\mathcal{M}_B$ such that
\begin{eqnarray}\label{eq242}
(\rho_\mathcal{E}^R)_{\alpha \beta} = \sum_{ \mu ,\nu} \left(
\mathcal{M}^A \otimes \mathcal{M}^B \right)_{\alpha \beta, \mu
\nu} (\rho_s^R)_{\mu \nu}.
\end{eqnarray}
   The
real-valued Mueller matrices $M_A$ and $M_B$ associated via Eq.
(\ref{eq24}) to
 $\mathcal{M}_A$ and $\mathcal{M}_B$, respectively,  can be
 written as
\begin{eqnarray}\label{eq250}
M_A = \left[ \begin{array}{cc}
  1 & \mathbf{0}^T  \\
  {\bf a} & A  \\
\end{array} \right], \qquad M_B = \left[ \begin{array}{cc}
  1 & \mathbf{0}^T  \\
  {\bf b} & B  \\
\end{array} \right],
\end{eqnarray}
where Eq. (\ref{eq30}) with $\mathbf{d}_A =
\mathbf{0}=\mathbf{d}_B$ and $M_{00}=1$ has been used, and
\begin{eqnarray}\label{eq260}
\mathbf{a} = \left[ \begin{array}{c}
  a_1 \\
  a_2 \\
  a_3
\end{array} \right], \qquad \mathbf{b} = \left[ \begin{array}{c}
  b_1 \\
  b_2 \\
  b_3
\end{array} \right],
\end{eqnarray}
are the polarizance vectors of $M_A$ and $M_B$, respectively. We
remember that the condition $\mathbf{d}_A =
\mathbf{d}_B=\mathbf{0}$ is a consequence of the fact that both
maps $\mathcal{E}_A$ and $\mathcal{E}_B$ are trace-preserving,
while the conditions $\mathbf{a} \neq \mathbf{0}$ and $\mathbf{b}
\neq \mathbf{0}$ reflect the non-unital nature of $\mathcal{E}_A$
and $\mathcal{E}_B$. With this notation we can rewrite Eqs.
(\ref{eq230}-\ref{eq240}) as
\begin{eqnarray}
\rho_\mathcal{E}^A  = \frac{1}{2} \sum_{\mu = 0}^3 a_\mu X_\mu ,
\label{eq270}\\
\rho_\mathcal{E}^B  = \frac{1}{2} \sum_{\nu = 0}^3 b_\nu X_\nu ,
\label{eq280}
\end{eqnarray}
where we have defined $a_0 = 1 = b_0$. Moreover, the output
two-qubit density matrix $\rho_\mathcal{E} = \mathcal{E}[\rho_s]$
can be decomposed into a real and an imaginary part as
 $\rho_\mathcal{E} = \rho_\mathcal{E}^\mathrm{Re} + i
\rho_\mathcal{E}^\mathrm{Im} $, where
\begin{eqnarray}
 \rho_\mathcal{E}^\mathrm{Re} = \frac{1}{4}\left[ \begin{array}{cccc}
 \alpha_{+}^+ & \beta_{+} &  \gamma_{+} &  \delta_{+}\\
 \beta_{+} &  \alpha_{-}^+ &   \delta_{-} &   \gamma_{-} \\
 \gamma_{+} &  \delta_{-} &  \alpha_{+}^- &   \beta_{-} \\
 \delta_{+} &  \gamma_{-}&   \beta_{-} & \alpha_{-}^-
\end{array} \right] ,\label{eq290}
\end{eqnarray}
and
\begin{eqnarray}\label{eq300}
 \rho_\mathcal{E}^\mathrm{Im} = \frac{1}{4}\left[ \begin{array}{cccc}
 0 & -\xi_{+} &  -\eta_{+} &  -\tau_{+}\\
 \xi_{+} & 0 &   -\tau_{-} &   -\eta_{-} \\
 \eta_{+} &  \tau_{-} &  0 &   -\xi_{-} \\
 \tau_{+} &  \eta_{-}&   \xi_{-} & 0
\end{array} \right] ,
\end{eqnarray}
with
\begin{eqnarray}\label{eq310}
\alpha_{\pm}^+ \equiv (1 + a_3) \pm [b_3(1+a_3) - C_{33}], \nonumber \\
\alpha_{\pm}^- \equiv (1 - a_3) \pm [b_3(1-a_3) + C_{33}],
\end{eqnarray}
and
\begin{eqnarray} \label{eq320}
\beta_{\pm} \equiv b_1 \pm (a_3 b_1 - C_{31}),\nonumber \\
\gamma_{\pm} \equiv a_1 \pm (a_1 b_3 - C_{13}), \nonumber \\
\delta_{\pm} \equiv a_1 b_1 - C_{11} \mp (a_2 b_2 - C_{22}),
\end{eqnarray}
and
\begin{eqnarray} \label{eq330}
\xi_{\pm} \equiv b_2 \pm (a_3 b_2 - C_{32}),\nonumber \\
\eta_{\pm} \equiv a_2 \pm (a_2 b_3 - C_{23}), \nonumber \\
\tau_{\pm} \equiv a_2 b_1 - C_{21} \pm (a_1 b_2 - C_{12}),
\end{eqnarray}
where $C_{ij} \equiv (AB^T)_{ij}, \, i,j \in \{1,2,3\}$.

At this point, our goal is to determine the two vectors
$\mathbf{a}, \, \mathbf{b}$ and the two $3 \times 3$ matrices $A,
\, B$ such that $ \rho_\mathcal{E}^\mathrm{Im} = 0$ and
\begin{eqnarray}\label{eq340}
\rho_\mathcal{E}^\mathrm{Re} = \rho_\mathrm{MEMS} = \left[
 \begin{array}{cccc}
   {g(p)}/{2} & 0 & 0 &  {p}/{2} \\
   0 & 1- g(p)& 0 & 0 \\
   0 & 0 & 0 & 0 \\
    {p}/{2} & 0 & 0 &  {g(p)}/{2} \
 \end{array}
 \right] ,
\end{eqnarray}
where
\begin{eqnarray}\label{eq350}
  g(p) = \left\{
\begin{array}{ccc}
  {2}/{3} \, , &  & 0  \leq p \leq {2}/{3}, \\\\
  p \, , &  &  {2}/{3} < p \leq 1 .
\end{array}
\right.
\end{eqnarray}
To this end, first we calculate $\mathbf{a}$ and $\mathbf{b}$ by
imposing:
\begin{eqnarray}
\rho_\mathcal{E}^A =\rho_\mathrm{MEMS}^A  = \left[
\begin{array}{cc}
  1-g(p)/2 & 0 \\
  0 & g(p)/2
\end{array} \right],
\label{eq360}\\
\rho_\mathcal{E}^B = \rho_\mathrm{MEMS}^B  = \left[
\begin{array}{cc}
  g(p)/2 & 0 \\
  0 & 1-g(p)/2
\end{array} \right],
\label{eq370}
\end{eqnarray}
respectively. Note that only fulfilling  Eqs.
(\ref{eq360}-\ref{eq370}), together with $
\rho_\mathcal{E}^\mathrm{Re} =  \rho_\mathrm{MEMS}$ and
$\rho_\mathcal{E}^\mathrm{Im} =  0$, will ensure the achievement
of \emph{true} MEMS states. It is surprising that in the current
literature the importance of this point is neglected. Thus, by
solving Eqs. (\ref{eq360}-\ref{eq370}) we obtain $a_1 = a_2 = 0$,
$a_3 = 1 - g(p)$, and $\mathbf{b} = -\mathbf{a}$, where Eqs.
(\ref{eq270}-\ref{eq280}) have been used. Then, after a little of
algebra, it is not difficult to find that a possible bi-local map
$\mathcal{E} = \mathcal{E}_A \otimes \mathcal{E}_B$ that generates
a solution $\rho_\mathcal{E}$ for the equation $\rho_\mathcal{E} =
\rho_\mathrm{MEMS}$, can be expressed as in Eqs.
(\ref{eq242}-\ref{eq250}) in terms of the two real-valued Mueller
matrices
\begin{equation}
\begin{array}{rcl}
M_A  & = &\left[
\begin{array}{cccc}
1 & 0 & 0 & 0 \\
0 &  \sqrt{p} & 0 & 0 \\
0 &  0 & \sqrt{p} & 0 \\
1-g(p) &  0 & 0 & g(p)
\end{array}
 \right], \label{eq380}
\\\\
M_B & =  &  \left[
\begin{array}{cccc}
1 &  0 & 0 & 0\\
0 &  -\sqrt{p} & 0 & 0 \\
0 &  0 & \sqrt{p} & 0 \\
g(p)-1 &  0 & 0 & -g(p)
\end{array}
 \right].
\end{array}
\end{equation}
It is easy to check that both $M_A$ and $M_B$ are physically
admissible Mueller matrices since the associated matrices $H_A$
and $H_B$ have the same spectrum made of non-negative eigenvalues
$\{ \lambda_\mu \}=\{ \lambda_0, \, \lambda_1, \,\lambda_2,
\,\lambda_3\}$. In particular:
\begin{eqnarray}
\{ \lambda_\mu \} =  \left\{ 0, \, 1-p, \, 0, \, 1+p \right\},
\quad \mathrm{for} \quad {2}/{3} < p \leq 1 . \label{eq384}
\end{eqnarray}
 and
\begin{eqnarray}
\{ \lambda_\mu \} =  \left\{  0, \frac{1}{3},  \frac{5 -
\sqrt{1+36 p}}{6},   \frac{5 + \sqrt{1+36 p}}{6} \right\},
\label{eq382}
\end{eqnarray}
for $0 \leq p \leq {2}/{3}$.
It is also easy to see that the map $\mathcal{E}$ can be
decomposed as in Eq. (\ref{eq210}) in a Kraus sum
  with $A_0 = A_2 = 0$,
\begin{eqnarray}\label{eq390}
 A_1 \sqrt{\lambda_1} = \left[
\begin{array}{cc}
  0 & \sqrt{1-p} \\
  0 & 0
\end{array}
\right], \quad A_3 \sqrt{\lambda_3} = \left[
\begin{array}{cc}
  1 & 0 \\
  0 & \sqrt{p}
\end{array}
\right],
\end{eqnarray}
and $B_0 = B_2 = 0$,
\begin{eqnarray}\label{eq400}
B_1 \sqrt{\lambda_1} =\left[
\begin{array}{cc}
  0 & 0 \\
  0 &  \sqrt{1-p}
\end{array}
\right], \quad B_3 \sqrt{\lambda_3} = \left[
\begin{array}{cc}
  0 & -\sqrt{p} \\
  1 & 0
\end{array}
\right],
\end{eqnarray}
for $  {2}/{3} < p \leq 1$. Analogously, for $0  \leq p \leq
{2}/{3}$ we have $A_0 =  0$,
\begin{eqnarray}
A_1 \sqrt{\lambda_1}  =  \left[
\begin{array}{cc}
  0 & 1/\sqrt{3} \\
  0 &  0
\end{array}
\right],  \label{eq410}
\end{eqnarray}
\begin{eqnarray}
A_2 \sqrt{\lambda_2}  =\left[
\begin{array}{cc}
  - \phi_- &0 \\
0 &   \psi_+
\end{array}
\right], \quad A_3 \sqrt{\lambda_3}  =\left[
\begin{array}{cc}
  \phi_+ &0 \\
0 &   \psi_-
\end{array}
\right], \label{eq420}
\end{eqnarray}
and $B_0 =  0$,
\begin{eqnarray}
B_1 \sqrt{\lambda_1}  =  \left[
\begin{array}{cc}
  0 & 0 \\
  0 &  1/\sqrt{3}
\end{array}
\right],  \label{eq440}
\end{eqnarray}
\begin{eqnarray}
B_2 \sqrt{\lambda_2}  =\left[
\begin{array}{cc}
 0 &  \psi_+ \\
\phi_-  & 0
\end{array}
\right], \quad B_3 \sqrt{\lambda_3} = \left[
\begin{array}{cc}
 0 &  -\psi_- \\
\phi_+  & 0
\end{array}
\right], \label{eq450}
\end{eqnarray}
where
\begin{eqnarray}
\phi_\pm \equiv \sqrt{ \frac{1}{2}\left( 1 \pm \frac{1 +
6p}{\sqrt{1 + 36p}}\right) }, \\  \label{eq460}
 \\ \nonumber
 \psi_\pm \equiv \sqrt{
\frac{1}{3}\left( 1 \pm \frac{1 - 9 p}{\sqrt{1 +
36p}}\right)}\label{eq462}.
\end{eqnarray}
Note that these coefficients satisfy the following relations:
\begin{eqnarray}\label{eq465}
\sum_{\mu=0}^3  \lambda_\mu A_\mu^\dagger A_\mu =
\phi_+^2 + \phi_-^2 = 1,\\
\sum_{\nu=0}^3  \lambda_\nu B_\nu^\dagger B_\nu =
 \frac{1}{3} + \psi_+^2 + \psi_-^2 = 1.
\end{eqnarray}
 A straightforward calculation shows that the single-qubit
maps $\mathcal{E}_A$ and $\mathcal{E}_B$ are trace-preserving but
\emph{not} unital, since
\begin{eqnarray}\label{eq470}
\sum_{\mu=0}^3  \lambda_\mu A_\mu A_\mu^\dagger = \left[
\begin{array}{cc}
 2 - g(p) & 0 \\
0  & g(p)
\end{array}
\right],
\end{eqnarray}
and
\begin{eqnarray}\label{eq480}
\sum_{\nu=0}^3 \lambda_\nu B_\nu B_\nu^\dagger = \left[
\begin{array}{cc}
 g(p) & 0 \\
0  & 2 - g(p)
\end{array}
\right].
\end{eqnarray}
At this point our task has been fully accomplished. However,
before concluding this subsection, we want to point out that both
maps $\mathcal{E}_A$ and $\mathcal{E}_B$ must depend on the
\emph{same} parameter $p$ in order to generate proper MEMS states.
This means that either a classical communication must be
established between $\mathcal{T}_A$ and $\mathcal{T}_B$ in order
to fix the same value of $p$ for both devices, or a classical
signal encoding the information about the value of $p$ must be
sent towards both  $\mathcal{T}_A$ and $\mathcal{T}_B$.
\subsubsection{Physical implementation}
Now we furnish  a straightforward physical implementation for the
quantum maps presented above. Up to now, several linear optical
schemes generating MEMS states were proposed and experimentally
tested. Kwiat and coworkers \cite{Peters} were the first to
achieve MEMS
 using photon pairs from spontaneous parametric down
conversion. Basically, they induced decoherence in SPDC pairs
initially prepared in a pure entangled state  by coupling
polarization and \emph{frequency} degrees of freedom of the
photons. At the same time, a somewhat different scheme was used by
De Martini and coworkers \cite{Barbieri} who instead used the
spatial degrees of freedom of SPDC photons to induce decoherence.
In such a scheme the use of spatial degrees of freedom of photons
 required  the
manipulation of not only the emitted SPDC photons, but also of the
pump beam.

In this subsection, we show that both single-qubit maps
$\mathcal{E}_A$ and $\mathcal{E}_B$ can be physically implemented
as linear optical networks \cite{Skaar} where \emph{polarization}
and \emph{spatial} modes of photons are suitably coupled, without
acting upon the pump beam.
 The basic building blocks of such networks are
polarizing beam splitters (\textsf{PBS}s), half-waveplates
(\textsf{HWP}s), and mirrors.
 Let $|i, N \rangle $ be a
single-photon basis, where the indices $i$ and $N $ label
polarization and spatial modes of the electromagnetic field,
respectively. We can also write $|i, N \rangle = \hat{a}_{i
N}^\dagger | 0 \rangle$ in terms of the annihilation operators
$\hat{a}_{i N}$ and the vacuum state $| 0 \rangle$. A polarizing
beam splitter distributes horizontal ($i=H$) and vertical ($i=V$)
polarization modes over two distinct spatial modes, say $N=n$ and
$N=m$, as follows:
\begin{equation}\label{eq490}
\begin{array}{lcl}
|H, n \rangle_\inp \rightarrow |H, n \rangle_\out & \mathrm{and}
&|V, n
\rangle_\inp  \rightarrow |V, m \rangle_\out, \\\\
 |H, m \rangle_\inp \rightarrow |H, m \rangle_\out & \mathrm{and} &
|V, m \rangle_\inp \rightarrow |V, n \rangle_\out,
\end{array}
\end{equation}
as illustrated in Fig. 3.
%
%%%%%%%%%%%%%%%%%%%%%%%%%%%%%%%%%%%%%%%%%%%%%%%%%%%%%%%%%%%%%%%%%%%%%%%%%%%%%%%%%%%%%%%%%%%
\begin{figure}[hb!]
\includegraphics[angle=0,width=5truecm]{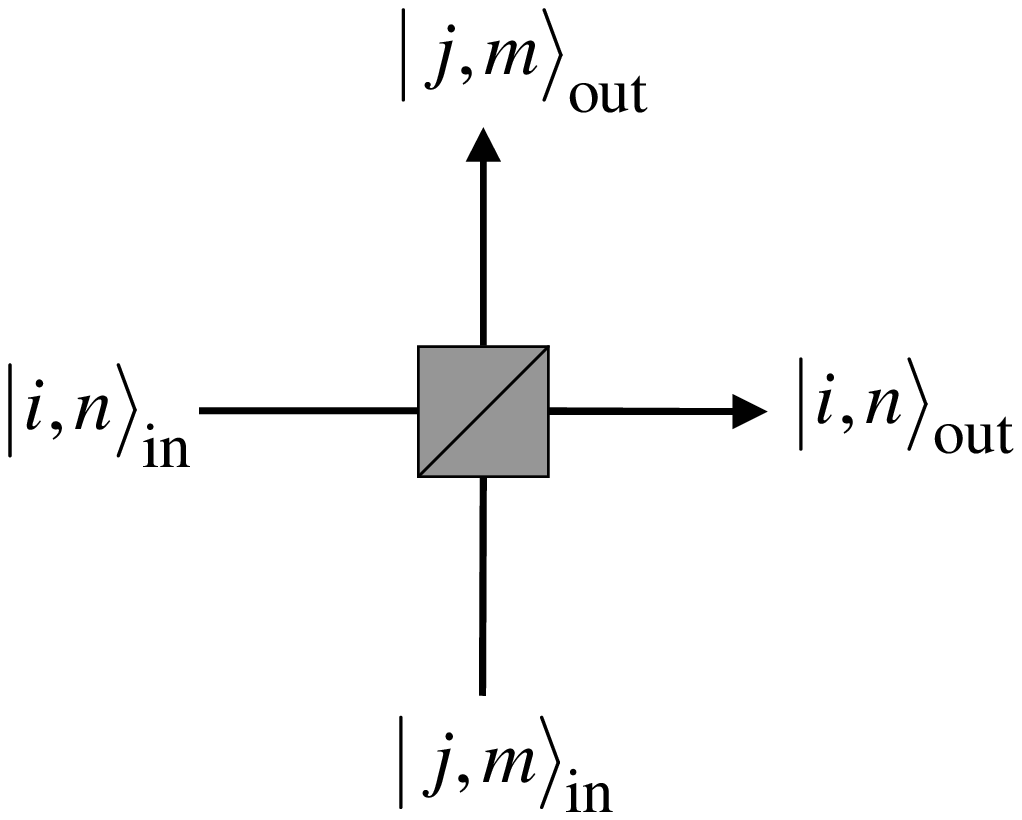}
\caption{\label{fig:3} The polarizing beam splitter couples
horizontal and vertical\emph{ polarization} modes ($i,j\in
\{H,V\}$), with two distinct \emph{spatial} modes  $N=n$ and $N=m$
of the  electromagnetic field.}
\end{figure}
%%%%%%%%%%%%%%%%%%%%%%%%%%%%%%%%%%%%%%%%%%%%%%%%%%%%%%%%%%%%%%%%%%%%%%%%%%%%%%%%%%%%%%%%%%%%%
%
A half-waveplate does not couple polarization and spatial modes of
the electromagnetic field and can be represented by a $2 \times 2$
Jones matrix $T_{HWP}(\theta)$ as
\begin{equation}\label{eq500}
T_{HWP}(\theta) = \left[
\begin{array}{cc}
  -\cos 2\theta &   -\sin 2\theta \\
    -\sin 2\theta &   \cos 2\theta
\end{array}
\right],
\end{equation}
where $\theta$ is the angle the optic axis makes with the
horizontal polarization. Two half-waveplates in series constitute
a polarization rotator represented by
$T_{R}(\theta)=T_{HWP}(\theta_0 + {\theta}/{2}
)T_{HWP}(\theta_0)$, where $\theta_0$ is an arbitrary angle and
\begin{equation}\label{eq510}
T_{R}(\theta) = \left[
\begin{array}{cc}
  \cos \theta &   -\sin \theta \\
    \sin \theta &   \cos \theta
\end{array}
\right].
\end{equation}
By combining these basic elements, composite devices may be built.
Figures 4 (a-b) show the structure of a horizontal (a), and
vertical (b) variable beam splitter, denoted \textsf{HVBS} and
\textsf{VVBS}, respectively. \textsf{HVBS} performs the following
transformation
%
%%%%%%%%%%%%%%%%%%%%%%%%%%%%%%%%%%%%%%%%%%%%%%%%%%%%%%%%%%%%%%%%%%%%%%%%%%%%%%%%%%%%%%%%%%%
\begin{figure}[!hb]
\includegraphics[angle=0,width=7truecm]{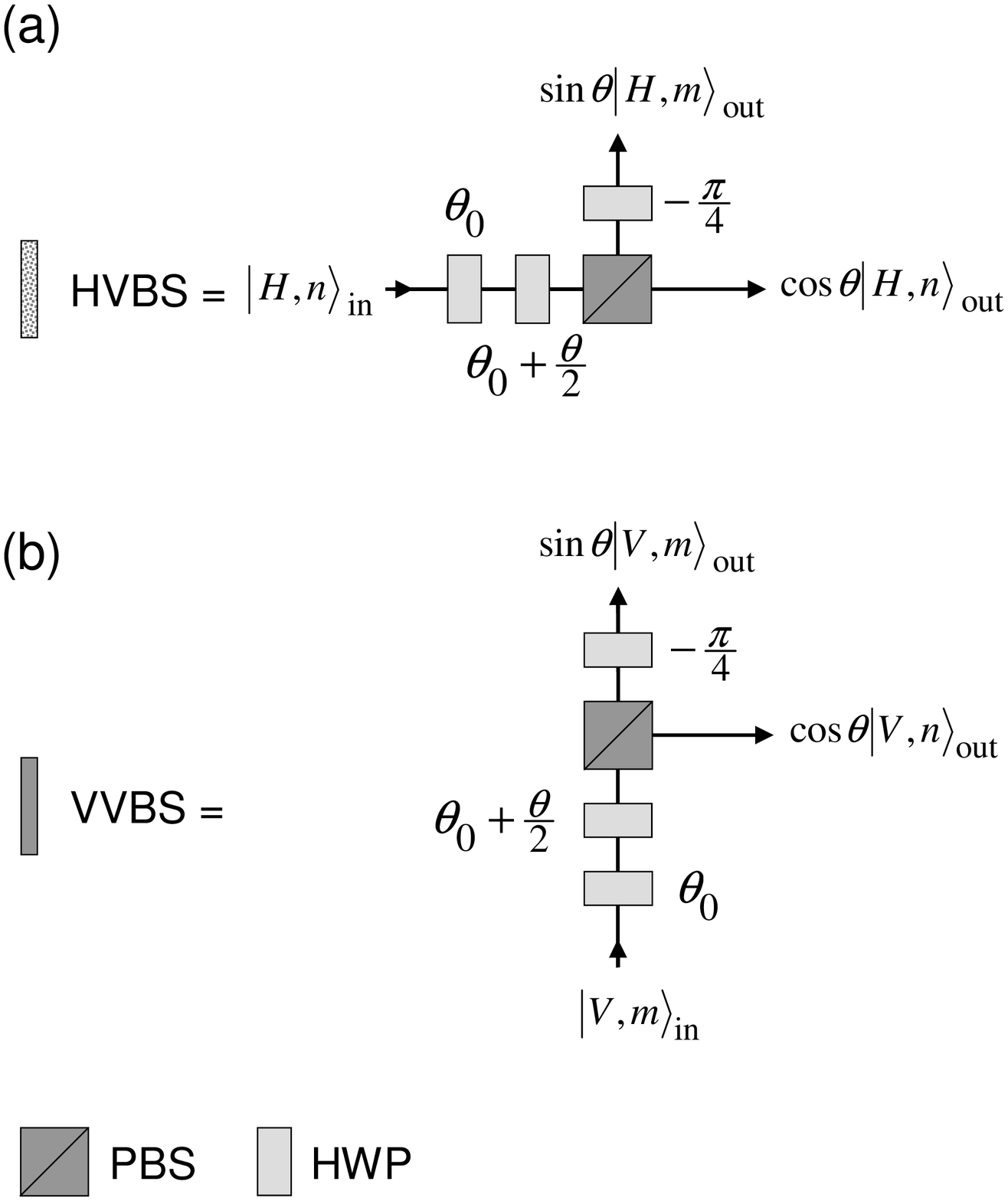}
\caption{\label{fig:4} The variable beam splitters \textsf{HVBS}
and \textsf{VVBS}.}
\end{figure}
%%%%%%%%%%%%%%%%%%%%%%%%%%%%%%%%%%%%%%%%%%%%%%%%%%%%%%%%%%%%%%%%%%%%%%%%%%%%%%%%%%%%%%%%%%%%%
%
%
\begin{equation}\label{eq520}
|H, n \rangle_\inp  \rightarrow  \cos \theta |H, n \rangle_\out +
\sin \theta|H, m \rangle_\out,
\end{equation}
while \textsf{VVBS} makes
\begin{equation}\label{eq530}
|V, m \rangle_\inp  \rightarrow  \cos \theta |V, n \rangle_\out +
\sin \theta|V, m \rangle_\out.
\end{equation}
At this point we have all the ingredients necessary  to built the
optical linear networks corresponding to our maps. We begin by
illustrating in detail  the optical network implementing
$\mathcal{E}_A$ (for $2/3 < p \leq 1$), which is shown in Fig. 5.
%
%%%%%%%%%%%%%%%%%%%%%%%%%%%%%%%%%%%%%%%%%%%%%%%%%%%%%%%%%%%%%%%%%%%%%%%%%%%%%%%%%%%%%%%%%%%
\begin{figure}[!ht]
\includegraphics[angle=0,width=4truecm]{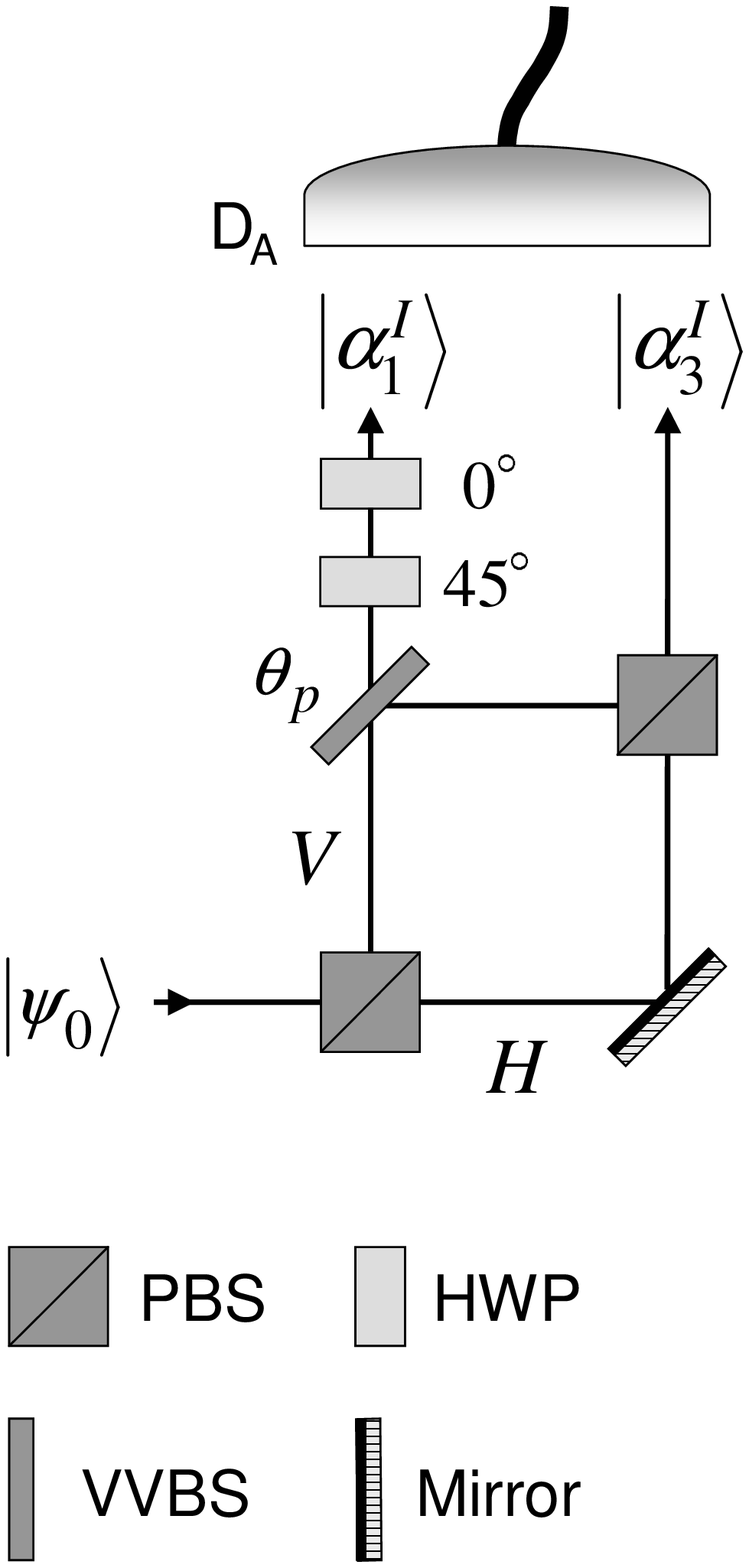}
\caption{\label{fig:5} Linear optical network implementing
$\mathcal{E}_A$ (for $2/3 < p \leq 1$), for MEMS I generation.}
\end{figure}
%%%%%%%%%%%%%%%%%%%%%%%%%%%%%%%%%%%%%%%%%%%%%%%%%%%%%%%%%%%%%%%%%%%%%%%%%%%%%%%%%%%%%%%%%%%%%
%
Let $| \psi_0 \rangle = a| H \rangle + b| V \rangle $ be the input
single-photon state entering the network. If we define the
\textsf{VVBS} angle
\begin{equation}\label{eq535}
\theta_p = \arccos \sqrt{p},
\end{equation}
then it is easy to obtain after a straightforward calculation:
\begin{eqnarray}
| \alpha_1^I \rangle = \sqrt{\lambda_1}A_1 | \psi_0 \rangle =
b\,\sqrt{1-p}\,| H \rangle, \label{eq537}
\end{eqnarray}
\begin{eqnarray} | \alpha_3^I \rangle = \sqrt{\lambda_3}A_3
| \psi_0 \rangle = a| H \rangle + b \,\sqrt{p}\,| V
\rangle. \label{eq540}
\end{eqnarray}
Since detector $\mathbf{\textsf{D}}_\mathbf{\textsf{A}}$ does not
distinguish spatial mode $1$ from spatial mode $2$,  the two
states $| \alpha_1^I \rangle$ and $| \alpha_3^I \rangle$, sum
incoherently and  the single-photon  output density matrix can be
written as $\rho_{\mathcal{E}_A} = | \alpha_1^I \rangle \langle
\alpha_1^I | +  | \alpha_3^I \rangle \langle  \alpha_3^I |$, where
\begin{eqnarray}\label{eq550}
\rho_{\mathcal{E}_A} = \left[
\begin{array}{cc}
  |a|^2 + |b|^2(1-p) &   a b^*\sqrt{p} \\
  a^* b\sqrt{p} &   p |b|^2
\end{array}
\right].
\end{eqnarray}
Of course, if we write the input density matrix as $\rho_0 =|
\psi_0 \rangle \langle \psi_0 |$, it is easy to see that
\begin{eqnarray}\label{eq560}
\rho_{\mathcal{E}_A} = \sum_{\mu=0}^3 \lambda_\mu  A_\mu \rho_0
A_\mu^\dagger.
\end{eqnarray}
where Eqs. (\ref{eq390}) have been used. Equation (\ref{eq560}),
together with Eq. (\ref{eq230}), proves the equivalence between
the quantum map $\mathcal{E}_A$ and the linear optical setup shown
in Fig. 5. Note that the Mach-Zehnder interferometers present in
Figs. 5 and 6 are balanced, that is their arms have the same
optical length.
In a similar manner, we can physically implement $\mathcal{E}_B$
(for $2/3 < p \leq 1$),  in the optical network shown in Fig. 6,
where we have defined
\begin{equation}
|\beta_1^I \rangle = \sqrt{\lambda_1}B_1 | \psi_0 \rangle =
b\,\sqrt{1-p}\,| V \rangle, \label{eq570}
\end{equation}
\begin{eqnarray} | \beta_3^I \rangle = \sqrt{\lambda_3}B_3
| \psi_0 \rangle = -b \,\sqrt{p} \,| H \rangle + a\,| V \rangle,
\label{eq580}
\end{eqnarray}
and, again, $\rho_{\mathcal{E}_B} = | \beta_1^I \rangle \langle
\beta_1^I | + | \beta_3^I \rangle \langle \beta_3^I |$.
%
%%%%%%%%%%%%%%%%%%%%%%%%%%%%%%%%%%%%%%%%%%%%%%%%%%%%%%%%%%%%%%%%%%%%%%%%%%%%%%%%%%%%%%%%%%%
\begin{figure}[!ht]
\includegraphics[angle=0,width=4truecm]{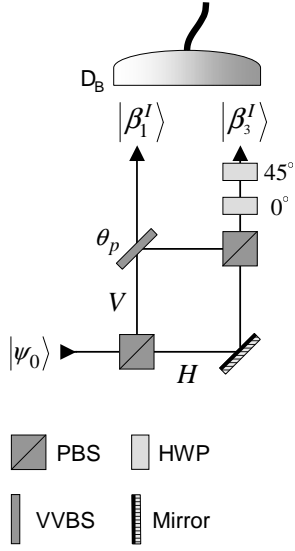}
\caption{\label{fig:6} Linear optical network implementing
$\mathcal{E}_B$ (for $2/3 < p \leq 1$), for MEMS I generation.}
\end{figure}
%%%%%%%%%%%%%%%%%%%%%%%%%%%%%%%%%%%%%%%%%%%%%%%%%%%%%%%%%%%%%%%%%%%%%%%%%%%%%%%%%%%%%%%%%%%%%
%

The optical networks necessary to realize quantum maps generating
MEMS II states are a bit more complicated. In order to illustrate
them we need to define the following two angles $\theta_{1/3}$ and
$\theta_{\psi}$ that determine the transmission amplitudes of two
\textsf{VVBS}s used in the MEMS II networks:
\begin{eqnarray}\label{eq590}
\theta_{1/3} = \arccos \sqrt{\frac{1}{3}}, \\
\theta_{\psi} = \arccos \left(\sqrt{\frac{3}{2}} \; \psi_+
\right).
\end{eqnarray}
In addition, a third angle $\theta_{\phi}$ determining the
transmission amplitudes of a \textsf{HVBS}, must be introduced:
\begin{eqnarray}\label{eq600}
\theta_{\phi} = \arccos \phi_+ .
\end{eqnarray}
Then, the map $\mathcal{E}_A$ (for $0 \leq p \leq 2/3$),  is
realized by the optical network shown in Fig. 7, where we have
defined
\begin{eqnarray}
| \alpha_2^{II} \rangle = \sqrt{\lambda_2}A_2 | \psi_0 \rangle =
-a\,\phi_-\,| H \rangle + b\,\psi_+\,| V \rangle, \label{eq610}
\end{eqnarray}
\begin{eqnarray} | \alpha_3^{II} \rangle = \sqrt{\lambda_3}A_3 | \psi_0 \rangle =
a\,\phi_+\,| H \rangle + b\,\psi_-\,| V \rangle, \label{eq620}
\end{eqnarray}
\begin{eqnarray} | \alpha_1^{II} \rangle = \sqrt{\lambda_1}A_1
| \psi_0 \rangle = \frac{b}{\sqrt{3}}| H \rangle . \label{eq630}
\end{eqnarray}
In this case, incoherent detection produces the output mixed state
 $\rho_{\mathcal{E}_A} = | \alpha_2^{II} \rangle \langle
\alpha_2^{II} | + | \alpha_3^{II} \rangle \langle  \alpha_3^{II}
|+ | \alpha_1^{II} \rangle \langle  \alpha_1^{II}|$.
%
%%%%%%%%%%%%%%%%%%%%%%%%%%%%%%%%%%%%%%%%%%%%%%%%%%%%%%%%%%%%%%%%%%%%%%%%%%%%%%%%%%%%%%%%%%%
\begin{figure}[!ht]
\includegraphics[angle=0,width=6truecm]{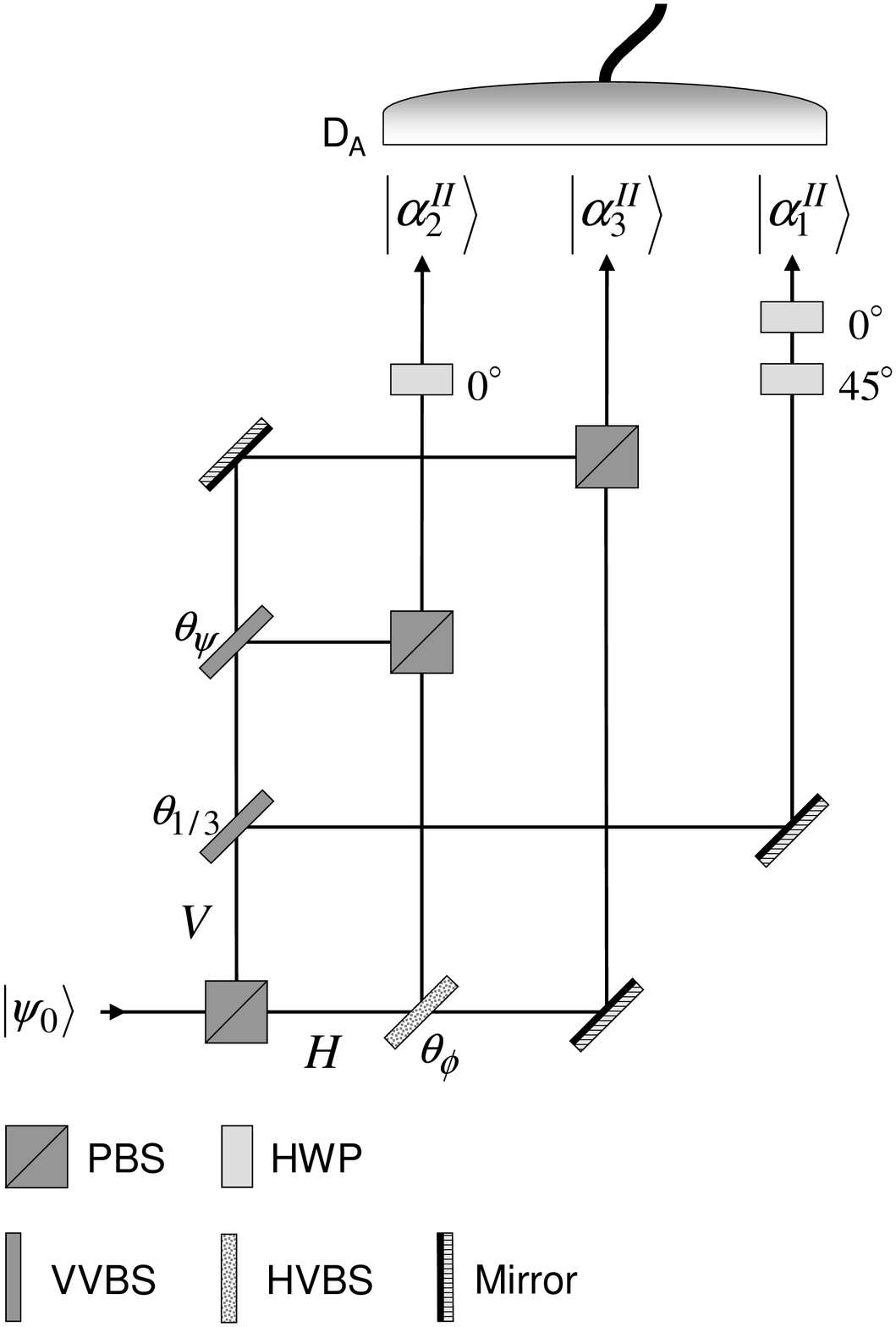}
\caption{\label{fig:7} Linear optical network implementing
$\mathcal{E}_A$ (for $ 0 \leq p \leq 2/3$), for MEMS II
generation. Each of the two Mach-Zehnder interferometers
constituting the network are balanced. }
\end{figure}
%%%%%%%%%%%%%%%%%%%%%%%%%%%%%%%%%%%%%%%%%%%%%%%%%%%%%%%%%%%%%%%%%%%%%%%%%%%%%%%%%%%%%%%%%%%%%
%
Finally, the map $\mathcal{E}_B$ (for $0 \leq p \leq 2/3$),  is
realized by the optical network shown in Fig. 8, where we have
defined
\begin{eqnarray}
| \beta_2^{II} \rangle = \sqrt{\lambda_2} B_2 | \psi_0 \rangle = b
\,\psi_+\,| H \rangle + a\,\phi_-\,| V \rangle, \label{eq640}
\end{eqnarray}
\begin{eqnarray} | \beta_3^{II} \rangle = \sqrt{\lambda_3} B_3 | \psi_0 \rangle =
-b\,\psi_-\,| H \rangle + a\,\phi_+\,| V \rangle, \label{eq650}
\end{eqnarray}
\begin{eqnarray} | \beta_1^{II} \rangle = \sqrt{\lambda_1} B_1
| \psi_0 \rangle = \frac{b}{\sqrt{3}}| V \rangle . \label{eq660}
\end{eqnarray}
%
%
%%%%%%%%%%%%%%%%%%%%%%%%%%%%%%%%%%%%%%%%%%%%%%%%%%%%%%%%%%%%%%%%%%%%%%%%%%%%%%%%%%%%%%%%%%%
\begin{figure}[!hr]
\includegraphics[angle=0,width=6truecm]{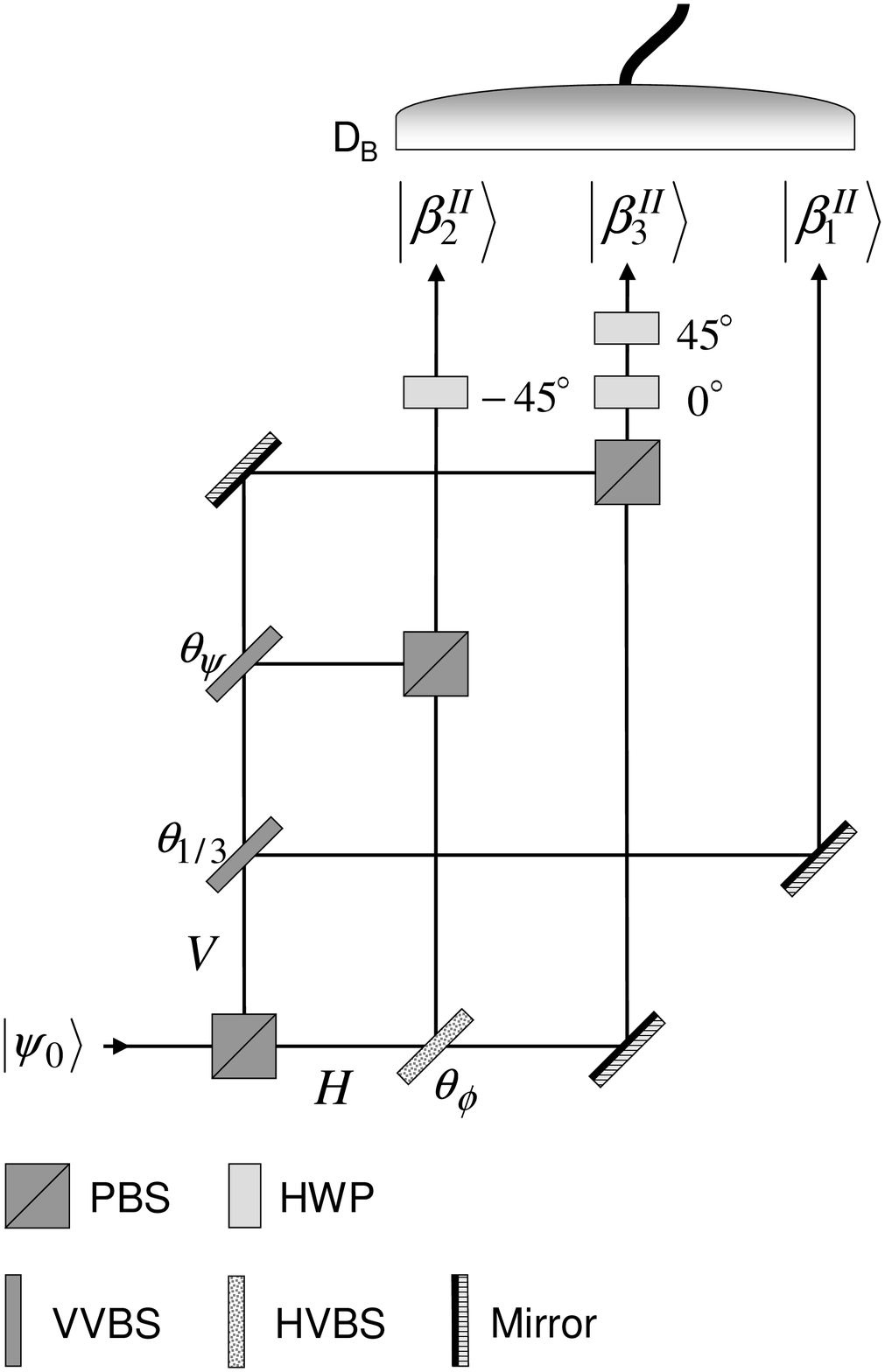}
\caption{\label{fig:8} Linear optical network implementing
$\mathcal{E}_B$ (for $ 0 \leq p \leq 2/3$), for MEMS II
generation. Each of the two Mach-Zehnder interferometers
constituting the network are balanced. }
\end{figure}
%%%%%%%%%%%%%%%%%%%%%%%%%%%%%%%%%%%%%%%%%%%%%%%%%%%%%%%%%%%%%%%%%%%%%%%%%%%%%%%%%%%%%%%%%%%%%
%
As before, now we have
 $\rho_{\mathcal{E}_B} = | \beta_2^{II} \rangle \langle
\beta_2^{II} | + | \beta_3^{II} \rangle \langle  \beta_3^{II} |+ |
\beta_1^{II} \rangle \langle  \beta_1^{II} |$.
\section{Summary and conclusions}
Classical polarization optics and quantum mechanics of two-level
systems are two different branches of physics that share the same
mathematical machinery.
In this paper we have described the analogies and connections
between these two subjects. In particular, after a review of the
 matrix formalism of classical polarization optics, we established
 the exact relation between one- and two-qubit quantum maps and
 classical description of linear optical processes. Finally, we
 successfully applied the formalism just developed, to two  cases
 of practical utility.

We believe that the present paper will be useful to both the
classical and the quantum optics community since it enlightens and
puts on a rigorous basis, the so-widely used relations between
classical polarization optics and quantum mechanics of qubits. A
particularly interesting aspect of our work is that we describe in
detail how dichroic devices (i.e., devices with
polarization-dependent losses), fit into this general scheme.
\begin{acknowledgments}
This project is supported by FOM.
\end{acknowledgments}


\begin{thebibliography}{41}
\bibitem[{Not()}]{NoteStokes}
\bibinfo{note}{Actually, our Stokes parameters $x_\alpha$ differ from the
  traditional ones $s_\alpha$ as given, e.g., in chapter $10$ of
  \cite{BornWolf}. However, the two sets of parameters are related by the
  simple relations: $s_0 = x_0, \, s_1 = x_3, \, s_2 = x_1, \, s_3 = -x_2$.}
\end{thebibliography}
\end{document}